\documentclass[a4paper,11pt]{article}
\pdfoutput=1 
\usepackage{jcappub} 

\usepackage{aasmacros}              
\usepackage{comment}
\usepackage{lscape}
\usepackage{verbatim}
\usepackage{comment}
\usepackage{array}
\usepackage{subcaption}
\usepackage{multirow}
\usepackage{graphicx}
\graphicspath{{graphics/}}
\usepackage{amsmath}
\usepackage{amssymb}
\usepackage{rotating} 
\usepackage{xspace}
\usepackage[dvipsnames]{xcolor}
\usepackage{booktabs}
\usepackage[normalem]{ulem}
\usepackage{adjustbox}
\usepackage{lipsum}  
\usepackage{caption}
\newcommand{\Msun}{{\rm M}_\odot}
\renewcommand{\epsilon}{\varepsilon}

\title{Directional dark matter signatures of the Large Magellanic Cloud}

\author[a,b]{Javier Reynoso-Cordova,}
\author[b,c]{Nassim Bozorgnia,}
\author[b]{and Marie-C\'ecile Piro}

\affiliation[a]{Istituto Nazionale di Fisica Nucleare, Sezione di Napoli, 
\\Complesso Universitario di Monte Sant'Angelo,
Via Cintia, 80126 Napoli, Italy}
\affiliation[b]{Department of Physics, University of Alberta,
CCIS 4-181, \\ Edmonton, Alberta T6G 2E1, Canada}
\affiliation[c]{Theoretical Physics Institute, University of Alberta, CCIS 4-181,\\ Edmonton, Alberta, Canada}

\emailAdd{javier.reynoso@na.infn.it}
\emailAdd{nbozorgnia@ualberta.ca}
\emailAdd{mariecci@ualberta.ca}

\abstract{The Large Magellanic Cloud (LMC), the most massive satellite of the Milky Way (MW), can significantly perturb the local dark matter (DM) distribution. We study its impact on directional DM detection using the Auriga cosmological simulations of a MW analogue hosting an LMC analogue. We find that the LMC induces strong anisotropies in directional recoil signals, driven primarily by the non-zero mean azimuthal velocity of the local DM distribution. The characteristic ring-like feature predicted in the Standard Halo Model (SHM) for heavy DM and low recoil energies is strongly distorted, producing an asymmetric recoil pattern concentrated at preferred azimuthal angles. Differences between recoil maps for the MW-LMC analogue and the SHM reach up to $\sim80\%$ near the signal maximum. These distortions significantly enhance directional discovery prospects, reducing the number of events required to reject isotropy by nearly a factor of five for a 100 GeV DM particle in a near-future CYGNUS-like experiment, and by even larger factors for heavier DM. Our results highlight the importance of the LMC for interpreting and optimizing future directional DM searches.}

\begin{document}
\maketitle
\flushbottom

\section{Introduction}
\label{sec:intro}

The nature of dark matter (DM) remains one of the  central open problems in particle physics and cosmology~\cite{Cirelli:2024ssz, Bozorgnia:2024pwk}. Direct detection experiments search for DM by measuring the small recoil energy deposited when DM particles scatter off nuclei or electrons in underground detectors~\cite{Goodman:1984dc,Lewin:1995rx,PhysRevD.33.3495}. Directional detection extends this strategy by measuring not only the recoil energy but also the direction of the recoiling nucleus. 

Because the Sun moves through the Galactic halo toward the constellation Cygnus, DM particles appear to arrive preferentially from that direction, producing a characteristic dipole anisotropy in the recoil distribution~\cite{Spergel:1987kx, Alenazi:2007sy, Ahlen:2009ev, Green:2010zm, Mayet:2016zxu,Vahsen:2021gnb,Baracchini:2023kyk}. The two directional DM signatures easiest to detect are departure from isotropy and the average recoil direction, while secondary signatures include ring-like~\cite{Bozorgnia:2011vc} and aberration features~\cite{Bozorgnia:2012eg} in the recoil rate. These directional signatures are extremely difficult for backgrounds to mimic and encode rich information about the underlying DM velocity distribution and particle physics interaction~\cite{Mayet:2016zxu,Vahsen:2021gnb,Baracchini:2023kyk, Bozorgnia:2011vc}. Directional measurements can therefore probe halo substructure, distinguish between interaction models, and enhance sensitivity to non-standard scenarios such as inelastic~\cite{Bozorgnia:2016qkh} or velocity-dependent scattering.

Beyond its role as a Galactic signature, directionality is also central to the future of direct detection experiments~\cite{Spergel:1987kx}. As experimental sensitivities approach the neutrino fog~\cite{OHare:2021utq}, nuclear recoils from Solar and atmospheric neutrinos become an irreducible background when only recoil energies are measured. Directional information, however, can separate a Galactic DM wind from neutrino induced recoils, whose angular distributions differ significantly in Galactic coordinates. This makes directional detection one of the few viable paths to strongly identify a DM signal in the presence of irreducible backgrounds. Experimentally, measuring the direction---and ideally the head-tail recognition with suitable angular resolution~\cite{Vahsen:2021gnb}---of keV scale nuclear recoils is challenging due to the extremely short track lengths involved~\cite{Mayet:2016zxu,Vahsen:2021gnb,Baracchini:2023kyk}. This has motivated the development of low-pressure gaseous time projection chambers such as DRIFT~\cite{Snowden-Ifft:1999reu,Burgos:2007zz}, MIMAC~\cite{Billard_2012,Santos_2018}, DMTPC~\cite{Ahlen:2010ub}, and NEWAGE~\cite{Miuchi:2010hn,Nakamura:2015iza,Shimada:2023vky}; low-pressure spherical proportional counter such as NEWS-G \cite{NEWS-G:2022kon, Coquillat:2025ahd}; nuclear emulsion detectors like NEWSdm~\cite{Alexandrov:2019gme,Golovatiuk:2021jcf}; anisotropic scintillators including NaI(Tl) and organic crystals~\cite{SHIMIZU2003347,Schuster:2016kpt}; and columnar recombination techniques~\cite{Nygren:2013nda, Mohlabeng:2015efa}. These technologies are converging toward scalable concepts such as the CYGNUS Galactic Recoil Observatory~\cite{Baracchini:2020btb,Vahsen:2020pzb, Lisotti:2024fco, Amaro:2022gub}, positioning directional detection as a realistic next generation strategy capable of probing both DM and neutrinos.

The predicted signals in direct detection experiments, both non-directional and directional, depend sensitively on the local DM phase space distribution. The most widely adopted model of the DM halo is the Standard Halo Model (SHM)~\cite{PhysRevD.33.3495}, in which the Milky Way (MW) halo is modeled as an isotropic, isothermal sphere with a Maxwell-Boltzmann velocity distribution in the Galactic  frame. Recent high-resolution cosmological simulations show that a Maxwellian velocity distribution generally provides a good description of the local DM velocity distribution in MW-like halos. However, substantial halo-to-halo scatter is observed among simulated galaxies, leading to large astrophysical uncertainties in the predicted local DM distribution and consequently the interpretation of direct detection data~\cite{Bozorgnia:2016ogo, Kelso:2016qqj, Sloane:2016kyi,  Bozorgnia:2017brl,  Bozorgnia:2019mjk, Poole-McKenzie:2020dbo, Lawrence:2022niq, Kuhlen:2013tra, Lacroix:2020lhn, Santos-Santos:2023ubx}.

One of the most important modifications to the SHM arises from the impact of the Large Magellanic Cloud (LMC) on the local DM phase space distribution. As the most massive satellite of the MW, the LMC strongly perturbs both the stellar and the DM halo of the MW~\cite{GaravitoCamargo:2021tcp, Garavito-Camargo:2020lqm, Conroy_2021, Petersen_2020,  Peterson_2020_MNRAS, Cunningham:2020nlo, Cavieres_2025}. Both idealized~\cite{Besla:2019xbx,Donaldson:2021byu} and cosmological~\cite{Smith-Orlik:2023kyl} simulations indicate that the recent infall of the LMC can significantly enhance the high speed tail of the local DM velocity distribution. This can shift direct detection exclusion
limits by several orders of magnitude towards smaller cross sections and DM masses~\cite{Besla:2019xbx, Smith-Orlik:2023kyl}, with especially pronounced effects for non-standard scenarios such as velocity-dependent interactions and inelastic DM~\cite{Reynoso-Cordova:2024lmc}. Including the effects of the LMC also extends the parameter space accessible to experiments searching for ultraheavy DM with masses well above the TeV scale~\cite{Bozorgnia:2025lsl}. 

Understanding how astrophysical perturbations induced by the LMC reshape directional recoil signatures is therefore essential for fully exploiting the discovery and diagnostic potential of future directional experiments. In this work, we use the Auriga cosmological magneto-hydrodynamical simulations~\cite{Grand:2016mgo} to investigate the impact of the LMC on the directional properties of the local DM flux in a simulated MW-LMC analogue. Building on our earlier analyses~\cite{Smith-Orlik:2023kyl, Reynoso-Cordova:2024lmc}, we quantify the resulting modifications to the directional recoil signal and examine their implications for the projected performance of a CYGNUS-like experiment.

The paper is structured as follows. In section~\ref{sec:sims} we present the details of the Auriga simulations and the MW-LMC analogue used in this work. In section~\ref{sec:vel} we present the  components of the DM velocity distribution in the Solar neighborhood. In section~\ref{sec:Radon_transform} we introduce the formalism for computing the directional event rate. In section~\ref{sec:experiments} we discuss the details of the directional experiments considered in this work. In section~\ref{sec:results} we present our results for the directional signatures of the LMC and perform a statistical analysis to specify the number of events required in directional experiments to reject isotropy. Finally, in section~\ref{sec:summary} we provide a brief discussion and our conclusions. In appendices~\ref{sec:appendix_numerical_int} and \ref{sec:lmc_sampling}, we provide additional materials relevant to this work.

\section{Simulations}
\label{sec:sims}

In this work, we analyze the simulated MW-LMC system originally identified  in ref.~\cite{Smith-Orlik:2023kyl}. The system is drawn from Auriga magneto-hydrodynamical simulations~\citep{Grand:2016mgo, Grand:2024}, which consists of a set of zoom-in simulations of isolated MW mass halos selected from a periodic volume of $100^3$~Mpc$^3$ (L100N1504) from the EAGLE project~\cite{Schaye:2014tpa, Crain:2015poa}. The simulations were carried out using the moving-mesh code Arepo~\citep{Springel:2009aa} and incorporate a galaxy formation subgrid model which includes star formation, black hole formation, metal cooling, active galactic nuclei and supernova feedback, and background UV/X-ray photoionisation radiation~\cite{Grand:2016mgo}. The  simulations successfully reproduce several observed properties of present day MW–mass galaxies, such as
stellar masses and sizes, rotation curves, star formation rates and metallicities. The simulations adopt the Planck-2015~\citep{Planck:2015fie} cosmological parameters: $\Omega_{m}=0.307$, $\Omega_{\rm bar}=0.048$, $H_0=67.77~{\rm km~s^{-1}~Mpc^{-1}}$. We adopt the standard resolution level (Level 4) of the Auriga simulations, corresponding to a DM particle mass of $m_{\rm DM} \sim 3\times 10^5~\Msun$, a baryonic particle mass of $m_b=5\times10^4~\Msun$, and a Plummer equivalent gravitational softening length of $\epsilon=370$~pc~\citep{Power:2002sw,Jenkins2013}.

The MW-LMC analogue considered in this work is the re-simulated halo 13 introduced in ref.~\cite{Smith-Orlik:2023kyl}, which corresponds to the Auriga 25 halo and its LMC analogue. This system was re-simulated with finer snapshots close to the LMC analogue's pericenter approach. The virial mass of the MW analogue is $1.2 \times 10^{12}~\Msun$, while the halo mass of LMC analogue at infall is $3.2 \times 10^{11}~\Msun$. Further details regarding the selection of this system and its properties can be found in ref.~\cite{Smith-Orlik:2023kyl}. We use the present day MW-LMC snapshot for this system, which is the snapshot closest to the present day separation of the observed MW and LMC system.  The speed of the LMC analogue with respect to the host MW analogue at this snapshot  is 317~km/s, while its distance to host is $\sim 50$~kpc, which closely match  the observed values. 

The position and velocity of the  Sun within the simulated halo are chosen to reproduce the observed Sun-LMC geometry, following the procedure described  in ref.~\cite{Smith-Orlik:2023kyl}. First, we identify orientations of the MW analogue's  stellar disk  that yield  the same angle with the LMC analogue's orbital plane as observed in the real system. For each allowed disk orientation, we then determine candidate Solar positions by matching the angles formed by the  LMC's orbital angular momentum with the Sun's position and velocity vectors in the simulation to their observed values. Finally, among these candidates we select the Solar position that provides the  closest match of the angles between the Sun's velocity vector and the LMC's position and velocities with their observed values.

The \emph{Solar region} is defined as the intersection of two volumes: a spherical shell extending from 6 to 10 kpc from the center of the MW analogue (with the Sun at a galactocentric distance of $\sim 8$~kpc) and a cone with an opening angle of $\pi/4$, whose axis points toward the best fit Solar position and whose vertex lies at the galactic center (see figure~2 of ref.~\cite{Smith-Orlik:2023kyl}). This definition ensures that the region remains sensitive to the chosen Solar position while still containing  several thousand DM particles, as discussed in ref.~\cite{Smith-Orlik:2023kyl}.

\section{Dark matter velocity distribution}
\label{sec:vel}

The local DM velocity distribution is an important input in the computation of expected event rates in directional DM experiments. In this section, we present the components of this distribution in the Solar region obtained from the MW-LMC analogue.

\begin{figure}[t]
    \centering
    \includegraphics[width=0.6\textwidth]{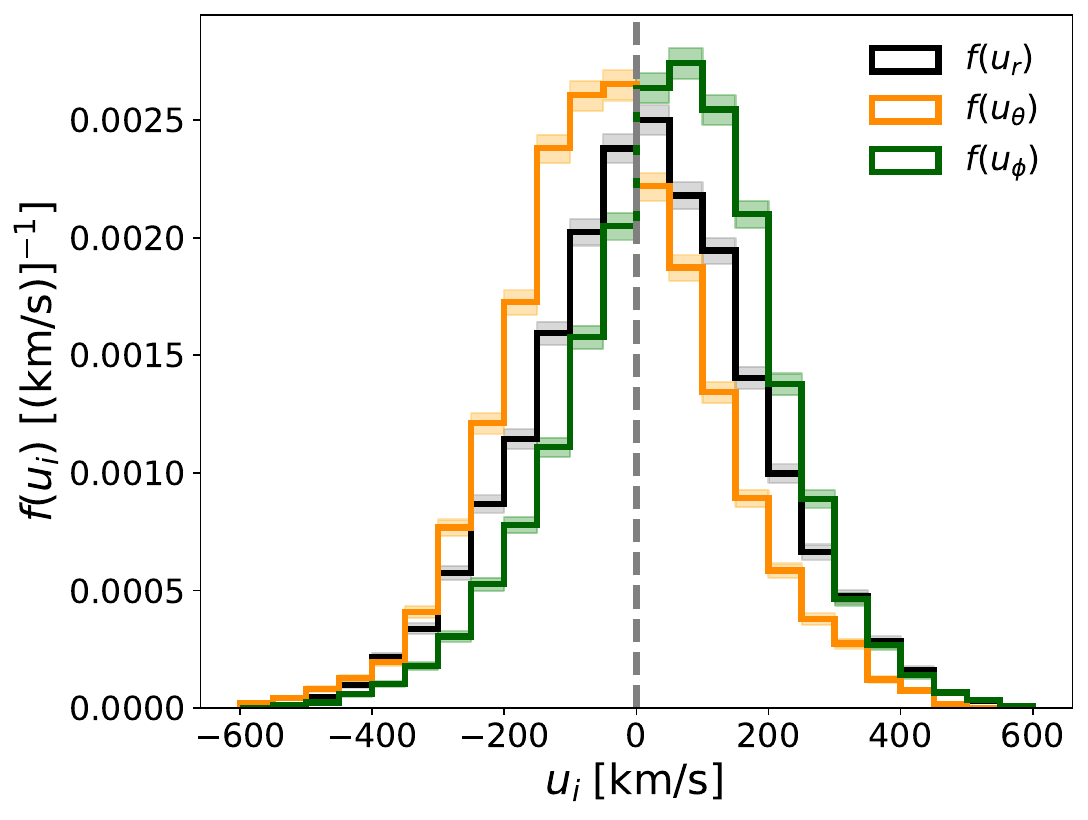}
    \caption{Radial (black), polar (orange), and azimuthal (green) components of the local DM velocity distribution in the Galactic frame for the simulated MW analogue. Shaded bands correspond to the $1\sigma$ Poisson errors in the distributions.} 
    \label{fig:velocity_components}
\end{figure}

We adopt a Galactocentric reference frame centered on the Galactic center, with the $x_{g}$-axis directed toward the Galactic center (such that the Sun is at $x_g\sim -8$~kpc), the $y_{g}$-axis oriented in the direction of the Galactic rotation, and the $z_{g}$-axis pointing toward the North Galactic Pole. The orientation of these axes is determined by the best fit position and velocity of the Sun that matches the observed Sun-LMC geometry, as discussed in section~\ref{sec:sims}. After transforming the positions and velocities of all simulation particles to the Galactic reference frame, we convert their Cartesian velocity components into  spherical components ($u_{r}$, $u_{\theta}$, $u_{\phi}$) relative to the Galactic center. Here, $u_{r}$ is the radial velocity component, $u_{\theta}$ is the polar velocity component (where $\theta$ is measured from the $z_g$-axis), and $u_{\phi}$ is the azimuthal velocity component. With this definition of the Galactic reference frame and the Solar position, a particle in the Solar region with negative $u_\phi$ would be co-rotating with the stellar disk.

Figure \ref{fig:velocity_components} shows the radial (black), polar (orange), and azimuthal (green) components of the local DM velocity distribution in the Galactic reference frame. The shaded bands specify the $1\sigma$ Poisson errors in the distributions. The distributions are normalized such that $\int du_i f(u_i)=1$, where $i=r, \theta, \phi$. Because in the SHM a local circular speed of 220~km/s is commonly adopted, the DM velocities in figure~\ref{fig:velocity_components} are scaled by $(220~{\rm km/s})/v_c$, where $v_c$ is the circular speed of the MW analogue at a Galactocentric distance of 8 kpc, computed from the enclosed mass. 

It is clear from figure~\ref{fig:velocity_components} that the local DM velocity distribution is anisotropic. While the radial and polar velocity distributions peak at velocities close to zero, the azimuthal velocity distribution peaks at $u_\phi \sim 75$~km/s. This indicates that DM particles in the Solar region are predominately counter-rotating relative to the stellar disk. This arises because the LMC analogue is moving in nearly the opposite direction of the Solar motion. Consequently, both DM particles originating from the LMC and MW DM particles boosted by the LMC tend to move towards the Sun, producing a counter-rotating azimuthal velocity component. 

We quantify this behavior with the velocity anisotropy parameter,
\begin{equation}
\beta = 1-\frac{\sigma_\theta^2+\sigma_\phi^2}{2\sigma_r^2}\,,
\end{equation}
where $\sigma_\theta$, $\sigma_\phi$, and $\sigma_r$ are  the polar, azimuthal, and radial velocity dispersions respectively, which we obtain directly from the local DM velocity distribution of the simulated MW analogue. The velocity distribution is isotropic  if $\sigma_\theta^2+\sigma_\phi^2=2\sigma_r^2$ and $\beta =0$, radially anisotropic if $\sigma_\theta^2+\sigma_\phi^2 \ll 2\sigma_r^2$ and $\beta \approx 1$, and tangentially anisotropic if $\sigma_\theta^2+\sigma_\phi^2 \gg 2\sigma_r^2$ and $\beta \to -\infty$. For the simulated MW analogue, we find $\beta=0.136$, which indicates a mildly radially anisotropic local DM velocity distribution.

\section{Directional event rate}
\label{sec:Radon_transform}

Directional DM experiments aim to measure both the recoil energy, $E_R$, and the 3D direction of a recoiling target nucleus after scattering by a DM particle. 

We consider a DM particle of mass $m_\chi$ elastically scattering with a target nucleus of mass $M_i$ in an underground detector. For the case of spin-independent scattering, the directional differential recoil rate is given by~\cite{Gondolo:2002np} 
\begin{equation}
    \frac{d^2 R}{dE_R d\Omega_q} = \frac{\rho_\chi\, \sigma_{\rm SI}}{4\pi m_\chi \mu_{\chi p}^2} \sum_i C_i\, A_i^2\, F^2_i(q)\,  \hat{f}_{\rm{det}}(v_{\rm min},\hat{\bf{q}})\,,
    \label{eq:dir_diff_spec}
\end{equation}
where $\rho_\chi$ is the DM density in the Solar neighborhood, $\sigma_{\rm SI}$ is the spin-independent DM-nucleon scattering cross section, and $\mu_{\chi p}$ is the reduced mass of the DM-nucleon system. The sum is over the nuclear species $i$ in the target, and $C_i$ and $A_i$ are the mass fraction and mass number of nuclide $i$, respectively. $F_i(q)$ is the nuclear form factor, for which we use the Helm~\cite{PhysRev.104.1466} form factor. $d\Omega_q$ is an infinitesimal solid angle around the recoil direction $\hat{\bf q}={\bf q}/q$, where ${\bf q}$ is the recoil momentum in the detector reference frame, and $q=|{\bf q}|$ is the magnitude of the recoil momentum. $\hat{f}_{\rm{det}}$ is the 3D Radon transform of the DM velocity distribution in the detector reference frame.

The minimum  speed, $v_{\rm min}$, required for a DM particle to transfer a recoil momentum $q$,  or equivalently deposit a recoil energy $E_R$ to the nucleus of mass $M$ is given by
\begin{equation}
v_{\rm min}= \frac{q}{2\mu_{\chi A}} =\sqrt{\frac{M E_R}{2\mu_{\chi A}^2}}\,,
\end{equation}
where $\mu_{\chi A}$ is the DM-nucleus reduced mass.

The Radon transform in the detector frame is defined as~\cite{Gondolo:2002np} 
\begin{equation}
    \hat{f}_{\rm det}(v_{\rm min},\hat{\mathbf{q}}) = \int \delta(\mathbf{v}\cdot\hat{\mathbf{q}} - v_{\rm min})\,f_{\rm det}(\mathbf{v})\,d^3v\, ,
    \label{eq:Radon_transform}
\end{equation}
where ${\bf v}$ is the velocity of DM with respect to the detector, and $f_{\rm{det}}(\mathbf{v})$ denotes the local DM velocity distribution in the detector rest frame normalized to one, such that $\int d^3 v f_{\rm det}({\bf v})=1$.

Dark matter distributions are typically defined in the Galactic rest frame. We can relate the Radon transform in the Galactic, $\hat{f}_{\rm{gal}}$, and detector, $\hat{f}_{\rm{det}}$, frames by~\cite{Gondolo:2002np}
\begin{equation}
    \hat{f}_{\rm{det}} (v_{\rm min},\hat{\bf{q}}) = \hat{f}_{\rm{gal}}(v_{\rm min} + \bf{V}_{\rm{lab}}\cdot \hat{\bf{q}},\hat{\bf{q}})\,,
    \label{eq:radon_lab}
\end{equation}
where $\bf{V}_{\rm{lab}}$ is the velocity of the detector with respect to the Galaxy and can be decomposed into four components: the Sun's circular velocity (or Local Standard of Rest (LSR) velocity), the peculiar velocity of the Sun with respect to the LSR, the velocity of Earth's revolution with respect to the Sun, and the velocity of Earth's rotation around itself. 

To compute the Radon transform in the detector frame, we  need to evaluate $\bf{V}_{\rm{lab}}\cdot \hat{\bf{q}}$. This requires expressing both $\bf{V}_{\rm{lab}}$ and $\hat{\bf{q}}$ in the same reference frame, and therefore transforming $\hat{\bf{q}}$ from the detector frame to the Galactic frame. The full transformation equations, as well as the resulting expression for $\bf{V}_{\rm{lab}}\cdot \hat{\bf{q}}$ are given in appendix A of ref.~\cite{Bozorgnia:2011vc}.

In the SHM, the local DM velocity distribution in the Galactic frame is modeled as an isotropic Maxwell-Boltzmann distribution with a peak speed equal to the local circular speed $v_c$, a 3D velocity dispersion of $\sigma_v=\sqrt{3/2}\,v_c$, and truncated at the Galactic escape speed  $v_{\rm esc}$. The Radon transform for the truncated Maxwellian distribution can be evaluated analytically. Applying the transformation in eq.~\eqref{eq:radon_lab} yields the Radon transform in the detector frame~\cite{Gondolo:2002np},
\begin{equation}
    \hat{f}_{\rm{det}}^{\rm MB}(v_{\rm min},\hat{\mathbf{q}}) = \frac{1}{N_{\rm esc}\left(2 \pi \sigma_v^2 \right)^{1/2} }\left[\exp\left(- \frac{(v_{\rm min} + \hat{\mathbf{q}}\cdot \mathbf{V}_{\rm{lab}})^2}{2\sigma_{v}^2}\right)-\exp\left(-\frac{v_{\rm esc}^2}{2\sigma_v^2
    }\right)\right]\,,
    \label{eq:radon_gaussian}
\end{equation}
if $v_{\rm min}+\hat{\bf q}\cdot {\bf V}_{\rm lab}<v_{\rm esc}$, and zero otherwise. Here
\begin{equation}
N_{\rm esc}={\rm erf}\left(\frac{v_{\rm esc}}{\sqrt{2}\sigma_v}\right)-\sqrt{\frac{2}{\pi}}\frac{v_{\rm esc}}{\sigma_v}\,\exp\left(-\frac{v_{\rm esc}^2}{2\sigma_v^2}\right).
\end{equation}

From eq.~\eqref{eq:radon_gaussian}, it can be seen that if $v_{\rm min} < |\mathbf{V}_{\rm{lab}}|$, the maximum of $\hat f_{\rm det}^{\rm MB}$ occurs when $-\hat{\mathbf{q}}\cdot \mathbf{V}_{\rm lab} = v_{\rm min}$, i.e.~the maximum will happen at an angle $\gamma$ between $\hat{\mathbf{q}}$ and $-\mathbf{V}_{\rm{lab}}$, forming a ring in recoil ${\bf q}$ space around the direction $-\mathbf{V}_{\rm{lab}}$~\cite{Bozorgnia:2011vc}.

\subsection{Radon transform from simulation data}
\label{sec:numerical_radon}

Next, we  compute the Radon transform of the local DM velocity distribution obtained directly from the simulations. The  DM speed distribution of the simulated MW analogue is boosted to higher speeds due to the impact of the LMC and  cannot be accurately modeled by a Maxwellian distribution. Therefore, the Radon transform, $\hat{f}_{\rm{gal}}(v_{\rm min} + \bf{V}_{\rm{lab}}\cdot \hat{\bf{q}},\hat{\bf{q}})$, must be computed numerically. 

Defining $w=v_{\rm min} + \bf{V}_{\rm{lab}}\cdot \hat{\bf{q}}$ and using the definition of the Radon transform, we  write
\begin{equation}
    \hat{f}_{\rm{gal}}(w,\hat{\mathbf{q}}) = \int \delta \left(\mathbf{u}\cdot\hat{\mathbf{q}} - w\right)f_{\rm{gal}}(\mathbf{u})\,d^3u\,,
    \label{eq:radon_gal_frame}
\end{equation}
where ${\bf u}={\bf v}+{\bf V}_{\rm lab}$ is the DM velocity in the Galactic rest frame.

The  DM velocity vector in Galactic coordinates is 
\begin{equation}
{\bf u}=u(\sin\theta \cos\phi, \sin\theta \sin\phi, \cos\theta)\, , 
\end{equation}
where $u=|{\bf u}|$, and $\theta$ and $\phi$ are the polar and azimuthal angles, respectively. The recoil direction $\hat{\bf q}$, defined in the detector reference frame,  can be expressed in the Galactic rest frame as
\begin{equation}
\hat{\mathbf{q}} = (\cos b \cos l, \cos b \sin l, \sin b)\, ,
\end{equation}
 with $l$ and $b$ the Galactic longitude and latitude, respectively.

To simplify the integral in eq.~\eqref{eq:radon_gal_frame}, we introduce a new coordinate system $(x',y',z')$ in which   $\hat{\mathbf{q}}$ is aligned with the $z'$-axis. The transformation from Galactic coordinates to this new reference frame is then given by the rotation matrix
\begin{equation}
\label{eq:rot_matrix}
R = \begin{pmatrix}
\sin b \cos l & \sin b \sin l & - \cos b \\
- \sin l & \cos l & 0 \\
\cos b \cos l & \cos b \sin l & \sin b
\end{pmatrix}.
\end{equation}

 By construction, in this  frame  $\mathbf{u}\cdot\hat{\mathbf{q}} = \mathbf{u} \cdot \hat{\mathbf{z}}' = u \cos \theta'$, where $\theta'$ is the polar angle of vector ${\bf u}$  in the primed  frame.
 
 Equation \eqref{eq:radon_gal_frame} then  becomes
  
\begin{equation}
\label{eq:radon_new_frame}
\hat{f}_{\rm gal}(w,\hat{\mathbf{q}}) = \int \delta(u \cos\theta' - w)f_{\rm{gal}}(\mathbf{u})\,u^2 du\, d\cos\theta' d\phi'\,,
\end{equation}
where $\phi'$ is the azimuthal angle of vector ${\bf u}$  in the primed  frame.

Using the properties of the Dirac delta, the integral reduces to 
\begin{equation}
\label{eq:integral_to_perform}
\hat{f}_{\rm gal}(w,\hat{\mathbf{q}}) = \int_{u\geq |w|} f_{\rm gal}(u,\lambda,\phi')\,u \,du\, d\phi'\,,
\end{equation}
where $\lambda \equiv \arccos(w/u)$. The condition $|\cos\lambda| \leq 1$ (equivalent to $u \geq |w|$) sets the lower limit of the velocity integral. For each recoil momentum direction $\hat{\bf q}$, the integrand $f_{\rm{gal}}(\mathbf{u})$ in eq.~\eqref{eq:integral_to_perform} must be evaluated at  ${\bf u}=(u,\lambda,\phi')$. 

In the primed frame, the DM velocity vector can be written as 
\[\mathbf{u} = u(\sin \theta' \cos \phi', \sin \theta' \sin \phi', \cos\theta'). \]
The relation between the Galactic and rotated frames depends on the direction $\hat{\mathbf{q}}(l, b)$, and is obtained by applying  the rotation $R$,
\begin{equation}
{\bf u'}=R\, {\bf u} \, ,
\end{equation}
with ${\bf u}$ expressed in Galactic coordinates. By equating the components of ${\bf u'}$ with those in the primed frame, we obtain the following relations between the angular coordinates, 
\begin{align}
\sin{\theta'}\cos{\phi'} &= \sin{b}\sin{\theta}\cos{(l-\phi)} - \cos{b}\cos{\theta} \, ,\nonumber\\
 \sin{\theta'}\sin{\phi'}  &= -\sin{\theta} \sin{(l-\phi)}\, ,\nonumber\\
\cos{\theta}' &= \cos{b}\sin{\theta}\cos{(l-\phi)} + \sin{b}\cos{\theta}\, .
\label{eq:angle_relations}
\end{align}

The DM velocity distribution extracted from the simulations is given in the Galactic reference frame. From this data, we construct a discrete three-dimensional velocity distribution, \(f_{\rm{gal}}(\mathbf{u})\), by binning the velocity components into a 3D histogram. We use a  bin size of 50~km/s for each Cartesian velocity component (\(u_{x}, u_{y}, u_{z}\)). To compute the Radon transform for a given recoil direction \(\hat{\mathbf{q}}(l,b)\) and minimum speed $v_{\rm min}$, we numerically evaluate the double integral in eq.~\eqref{eq:integral_to_perform}. For each point in the \((u, \lambda, \phi')\) integration grid, we use the relations in eq.~\eqref{eq:angle_relations} to find the corresponding  velocity vector in the Galactic frame and interpolate the binned distribution \(f_{\rm{gal}}\) to obtain the value of the integrand.

We implement the numerical integration using standard Python packages, with the angular integration over \(\phi'\) performed on a grid following the HEALPix convention \cite{Gorski:2004by} as implemented in the \texttt{healpy} library. We have conducted extensive tests to verify the numerical convergence and accuracy of our method. These tests explore the sensitivity of the results to the size of the velocity bins in the histogram, the number of integration points in \(u\) and \(\phi'\), the angular resolution (Nside parameter) of the HEALPix grid, and the intrinsic resolution of the simulation data. As a validation step, we have compared the numerical Radon transform of a Maxwellian velocity distribution generated from mock data with the analytical result from eq.~\eqref{eq:radon_gaussian}. We find a maximum relative difference of $\sim 7\%$, which remains stable when varying the sample size, bin size, and angular resolution, confirming that our numerical framework is robust within these parameter choices. A detailed discussion of these convergence tests is presented in appendix \ref{sec:appendix_numerical_int}, while appendix \ref{sec:lmc_sampling} assesses the stability of our results with respect to the finite resolution of the simulation sample.

\section{Directional experiments}
\label{sec:experiments}

In this work, we consider the near‑future CYGNUS‑like detector operated with a gaseous mixture at atmospheric pressure and room temperature, located at the Laboratori
Nazionali del Gran Sasso (LNGS). The corresponding detector specifications are listed in table~\ref{tab:cygnus_gransasso}.

\begin{table}[htbp]
\centering
\begin{tabular}{ll}
\hline
\textbf{Parameter} & \textbf{Value} \\
\hline
Targets & He:F$_4$ (60–40) \\ 
Energy Range & $[0.25\text{--}10]\,\mathrm{keV}$ \\
Exposure & $3.65 \times 10^{5}$  kg~day \\
Location & LNGS Gran Sasso \\
Coordinates & $(42.45^\circ,\ 13.57^\circ)$ \\
\hline
\end{tabular}
\caption{Specifications of the CYGNUS-like detector at Gran Sasso.}
\label{tab:cygnus_gransasso}
\end{table}

We have also verified our results for various underground laboratories and locations, as well as for different target materials, as summarized in table~\ref{tab:other_exp}. Additionally, we have tested for hypothetical labs located at the North/South Poles (90$^\circ$N/S) and at the Equator (0$^\circ$ latitude) for He:SF$_6$ (60–40) target. The impact of the LMC on the directional DM signatures is found to be very similar across all of these configurations. We therefore show our results for only  the CYGNUS-like experiment at Gran Sasso.

\begin{table}[htbp]
\centering
\begin{tabular}{lll}
\hline
\textbf{Experiment} & \textbf{Target} & \textbf{Lab Location}\\
\hline
CYGNUS-ANDES & He:SF$_6$ (60–40) & ANDES, Argentina \\ 
CYGNUS-Boulby & He:SF$_6$ (60–40) & Boulby, UK \\
CYGNUS-GranSasso & He:SF$_6$ (60–40) & LNGS, Italy \\
CYGNUS-Kamioka & He:SF$_6$ (60–40) & Kamioka, Japan \\
CYGNUS-Stawell & CS$_2$:O$_2$ (70–30) & Stawell, Australia \\
CYGNUS-SURF & He:SF$_6$ (60–40) & SURF, USA \\
MIMAC & CF$_4$:CHF$_3$:C$_4$H$_{10}$ (70–28–2) & Modane, France \\
NEWS-G & He:CH$_4$ (90–10) & SNOLAB, Canada\\
\hline
\end{tabular}
\caption{Directional detector configurations and locations.}
\label{tab:other_exp}
\end{table}

\section{Results}
\label{sec:results}

In  section~\ref{sec:results_rate}, we present the results of the directional differential recoil rate in a CYGNUS-like experiment for the local DM velocity distribution extracted from the simulated MW-LMC analogue and compare it to the SHM. In section~\ref{sec:statistical_analysis}, we perform a statistical analysis to find the number of events required to reject isotropy, when the impact of the LMC is included.

\subsection{Directional signatures}
\label{sec:results_rate}

The directional recoil rate in  eq.~\eqref{eq:dir_diff_spec} can be written, using  eq.~\eqref{eq:radon_lab}, as
\begin{equation}
    \frac{d^2 R}{dE_R d\Omega_q} = \left(0.048\, \frac{\rm{events}}{\rm{ton}\cdot \rm{yr} \cdot \rm{keV}\cdot\rm{sr}}\right) \frac{\rho_{0.3}\, \sigma_{48}}{4\pi m_\chi \mu_{\chi p}^2} \sum_i C_i\, A_i^2\, F^2_i(q)\,  \hat{f}_{\rm{gal}}(v_{\rm min} + \bf{V}_{\rm{lab}}\cdot \hat{\bf{q}},\hat{\bf{q}})\,,
\end{equation}
where $\rho_{0.3}=\rho_\chi/(0.3~\rm{GeV}\,\rm{cm}^{-3})$ and  $\sigma_{48}=\sigma_{\rm SI}/(10^{-48}\,\rm{cm}^2)$.

We consider the He:F$_4$ (60-40) target at Gran Sasso and two choices for the nuclear recoil energy near the experimental threshold: an ideal target value of $E_R=0.25$~keV and a more feasible threshold of $E_R=3$~keV~\cite{Vahsen:2020pzb, Vahsen:2021gnb, Mazzitelli:2023tdr}. We adopt a local DM density of $\rho_\chi=0.3\, \rm{GeV}\,\rm{cm}^{-3}$ and a DM–nucleon cross section of $\sigma_{\rm SI}=10^{-48} \; \rm{cm}^2$, compatible with direct detection constraints~\cite{Baxter:2021pqo, LZ:2024zvo}. We assume a date of June 2, 2026 in order to compute the laboratory's velocity and the recoil rate.

\begin{figure}[t]
    \centering
    \begin{subfigure}{0.47\textwidth}
        \centering
        \includegraphics[width=\linewidth]{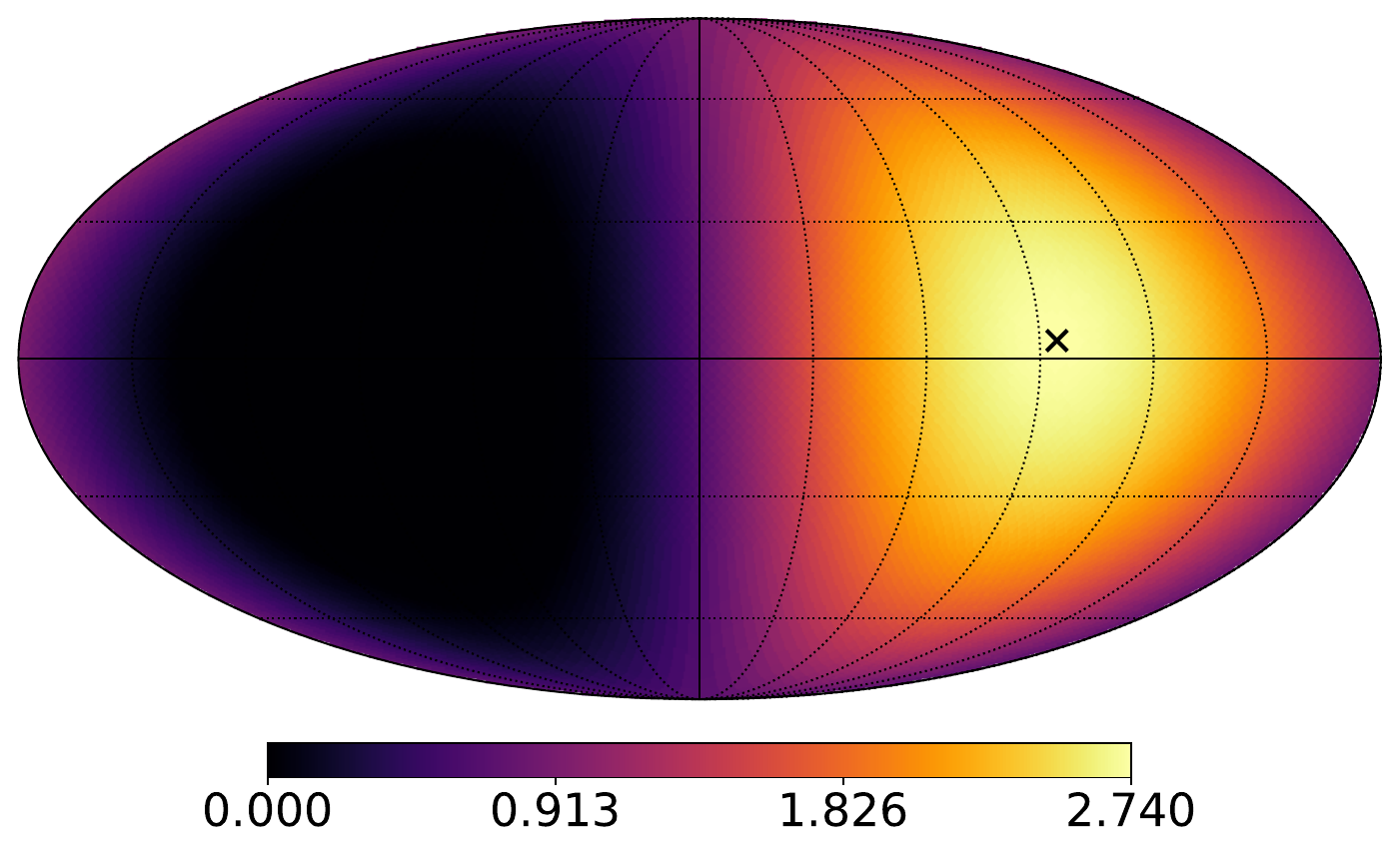}
    \end{subfigure}
     \hspace{0.4em} 
    \begin{subfigure}{0.47\textwidth}
        \centering
        \includegraphics[width=\linewidth]{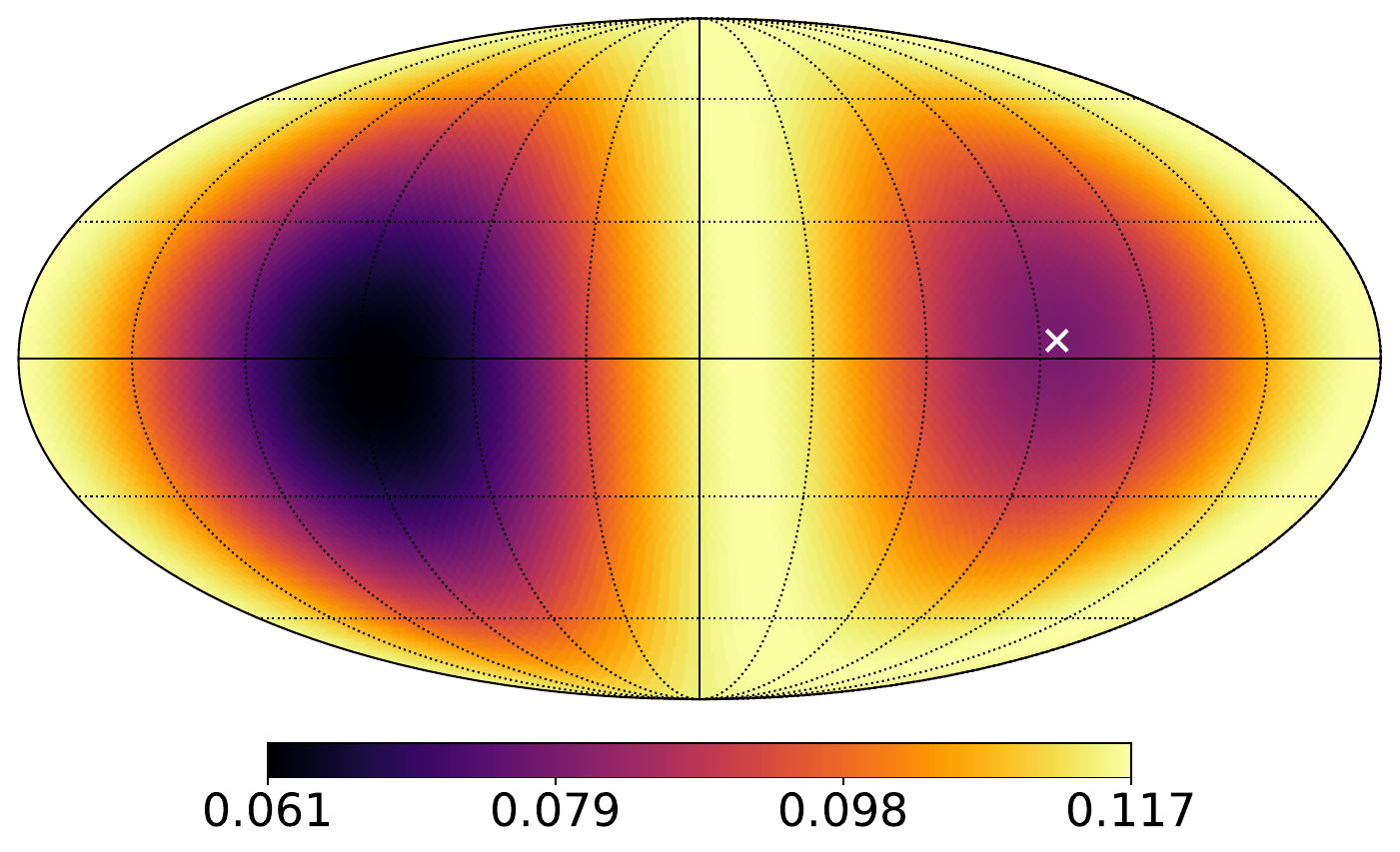}
    \end{subfigure}
    \caption{Mollweide equal-area projection maps of the celestial sphere in Galactic coordinates showing the directional differential recoil rate in a He:F$_4$ (60-40) target at Gran Sasso assuming the SHM Maxwellian DM velocity distribution. In the left panel we consider a DM particle mass of $m_\chi=5$~GeV and a nuclear recoil energy $E_R=3$~keV, while the right panel shows the case of  $m_\chi=100$~GeV and $E_R=0.25$~keV. The color bars show the values of the directional rate in units of $10^{-4}(\rho_{0.3}\sigma_{48}/\rm{ton}\cdot \rm{yr} \cdot \rm{keV} \cdot \rm{sr})$. The direction of $-{\bf V}_{\rm lab}$ is shown by a black and white cross in the left and right panels, respectively.}
    \label{fig:maps_gaussian_gran_sasso}
\end{figure}

Figure~\ref{fig:maps_gaussian_gran_sasso} shows Mollweide equal-area projection maps of the directional differential recoil rate in Galactic coordinates assuming the SHM Maxwellian DM velocity distribution (with Radon transform given in eq.~\eqref{eq:radon_gaussian}), 
with $v_c=220$~km/s 
and $v_{\rm{esc}} = 544\; \rm{km}/\rm{s}$. 
In the left panel,  
we assume a DM mass of $m_\chi =5$~GeV and a recoil energy of $E_R=3$~keV. This corresponds to $v_{\rm{min}}=332$~km/s and 396~km/s for the He and F target nuclei, respectively. Since $v_{\rm min}>|{\bf V}_{\rm lab}|$, the maximum recoil rate occurs in the direction of the average DM velocity, $-\mathbf{V}_{\rm{lab}}$, which is shown by a black cross on the map. In the right panel of the figure, we assume $m_\chi=100$~GeV and $E_R=0.25$~keV, resulting in $v_{\rm{min}}=57$~km/s and 30~km/s for He and F, respectively. Since in this case $v_{\rm min} < |\mathbf{V}_{\rm{lab}}|$, a ring of maximum recoil rate appears around the direction of $-\mathbf{V}_{\rm{lab}}$, which is shown by the white cross on the map.

\begin{figure}[t]
    \centering
    \begin{subfigure}{0.47\textwidth}
        \centering
        \includegraphics[width=\linewidth]{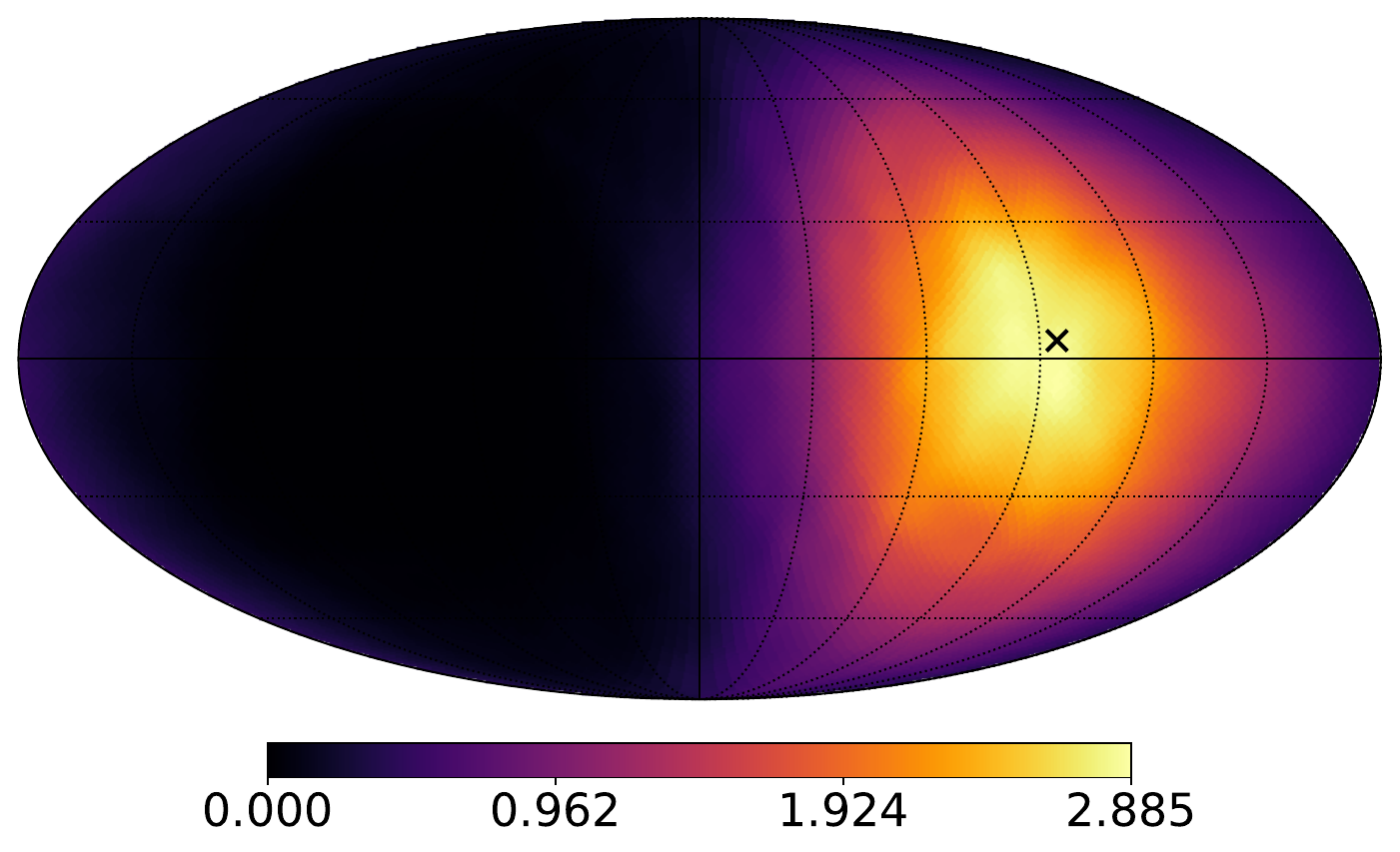}
    \end{subfigure}
     \hspace{0.4em} 
    \begin{subfigure}{0.47\textwidth}
        \centering
        \includegraphics[width=\linewidth]{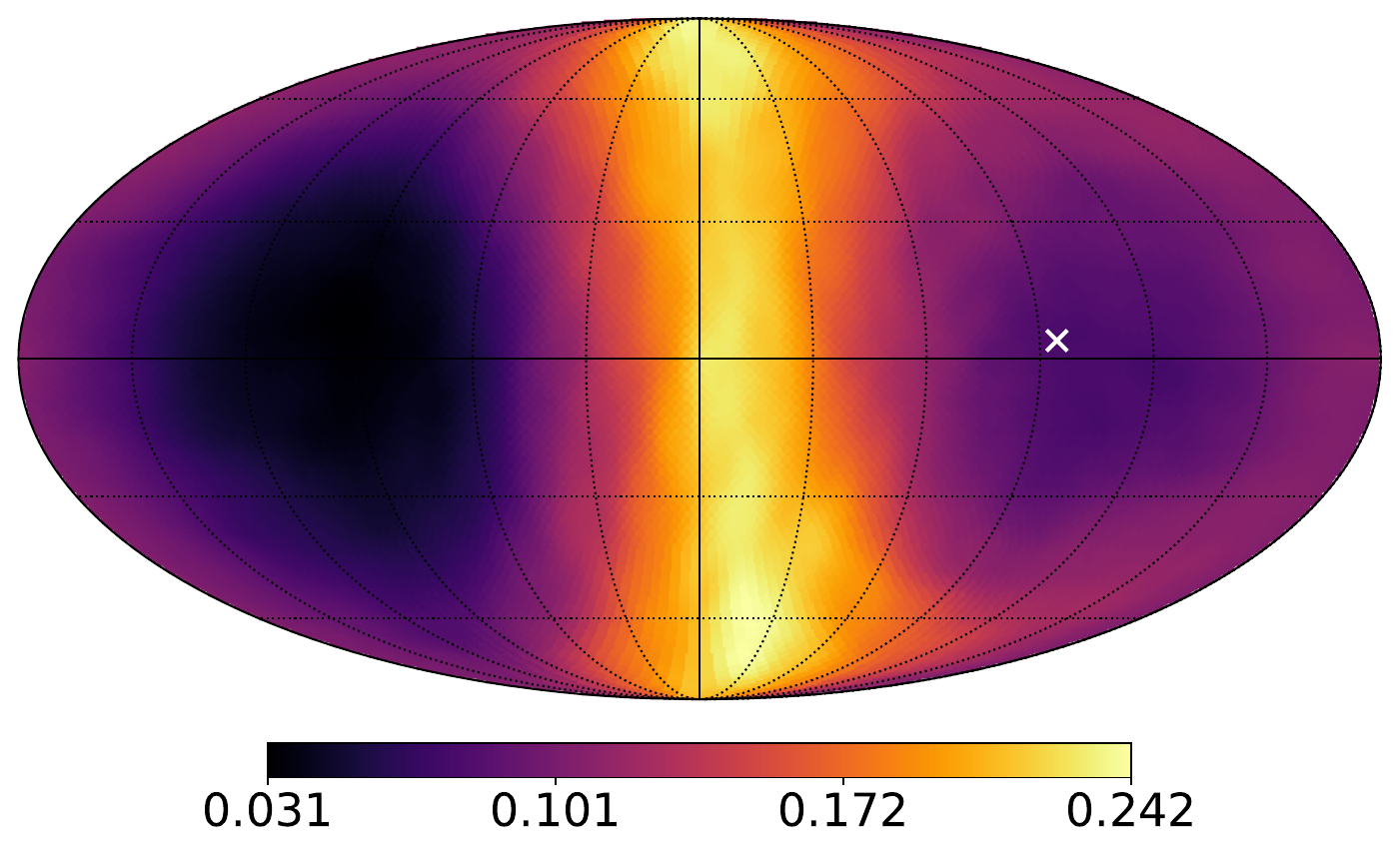}
    \end{subfigure}
    \caption{Same as figure~\ref{fig:maps_gaussian_gran_sasso}, but for the local DM velocity distribution extracted from the simulated MW-LMC analogue.}
    \label{fig:maps_lmc_gran_sasso}
\end{figure}

Figure~\ref{fig:maps_lmc_gran_sasso} shows the Mollweide map of the directional recoil rate for the same target nuclei and parameters used in figure~\ref{fig:maps_gaussian_gran_sasso}, but assuming the local DM velocity distribution extracted from the MW-LMC analogue. In the left panel of the figure, we show again the case of $m_\chi=5$~GeV and $E_R=3$~keV. As expected, the maximum recoil rate is in the direction of $-\mathbf{V}_{\rm{lab}}$, shown by a black cross in the map. This can be understood from the Sun-LMC geometry. The LMC is predominately moving in the opposite direction of the Sun. As a result, the high speed DM particles of the LMC and the DM particles of the MW boosted by the LMC are moving in the direction of $-{\bf V}_{\rm lab}$. This results in the maximum recoil rate to still occur in the direction of $-{\bf V}_{\rm lab}$ for the simulated MW-LMC analogue, when $v_{\rm min}>|{\bf V}_{\rm lab}|$.

In the right panel of figure~\ref{fig:maps_lmc_gran_sasso}, we show the case of $m_\chi =100$~GeV and $E_R=0.25$~keV. In this case, we observe a clear difference with respect to the recoil map of the Maxwellian distribution (shown in the right panel of figure~\ref{fig:maps_gaussian_gran_sasso}), as the ring-like signature is  replaced by a concentration of the maximum recoil rate at specific azimuthal angles around $-\mathbf{V}_{\rm{lab}}$. This distinct directional signature is primarily driven by the non-zero mean of the azimuthal velocity distribution induced by the LMC. We have checked that a directional recoil map produced assuming a triaxial Gaussian velocity distribution with the mean velocities fixed to the values obtained from the MW-LMC analogue shows generally similar features as those seen in the right panel of figure~\ref{fig:maps_lmc_gran_sasso}. 

To further quantify the effect of the LMC on the directional differential recoil rate, we compute the relative difference between the directional recoil map  for the SHM Maxwellian velocity distribution and that for the DM velocity distribution extracted from the simulated MW-LMC analogue. This relative difference is defined as
\begin{equation}
    \epsilon(\hat{\bf q}) = \left|1 - \left(\frac{d^2R}{d E_R d\Omega_q}\right)_{\rm{MW-LMC}}/\left(\frac{d^2R}{d E_R d\Omega_q}\right)_{\rm{SHM}}\right|\,.
    \label{eq:rel_diff}
\end{equation}

\begin{figure}[t]
    \centering
    \begin{subfigure}{0.47\textwidth}
        \centering
        \includegraphics[width=\linewidth]{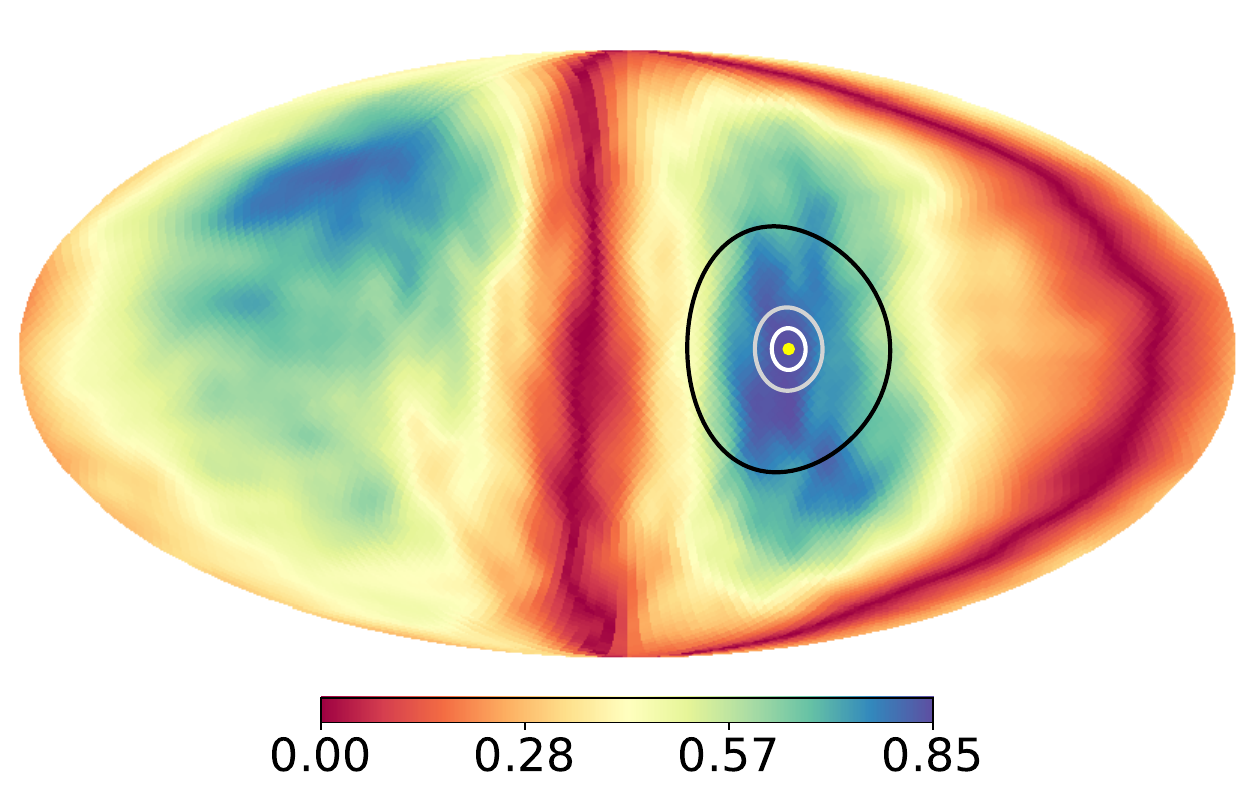}
    \end{subfigure}
    \hspace{0.4em}
     \begin{subfigure}{0.47\textwidth}
        \centering
        \includegraphics[width=\linewidth]{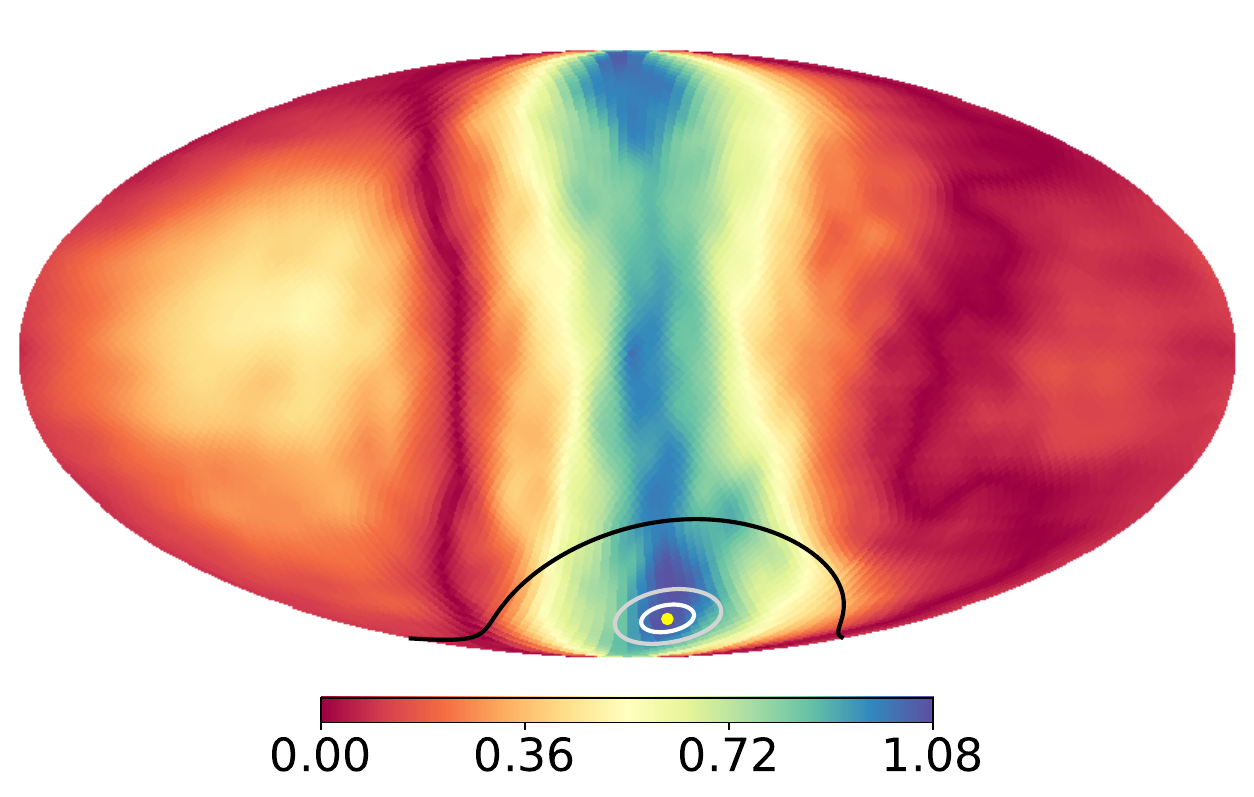}
    \end{subfigure}
    \caption{Absolute relative difference, $\epsilon(\hat{\bf q})$ (eq.~\eqref{eq:rel_diff}), between the directional recoil rate in a He:F$_4$ (60-40) target at Gran Sasso  computed assuming the SHM Maxwellian velocity distribution and  the velocity distribution extracted from the simulated MW-LMC analogue. In the left panel we consider a DM particle mass of $m_\chi=5$~GeV and a nuclear recoil energy $E_R=3$~keV, while the right panel shows the case of  $m_\chi=100$~GeV and $E_R=0.25$~keV.   
    The white, gray and black circles represent the angular regions $5^\circ$, $10^\circ$ and $30^\circ$, respectively, from the direction where the recoil rate computed assuming  the velocity distribution extracted from the simulated MW-LMC analogue is maximum (indicated with a yellow centered dot).
    }
    \label{fig:rel_diff}
\end{figure}

In figure~\ref{fig:rel_diff} we show the relative difference $\epsilon(\hat{\bf q})$, between the maps shown in figures~\ref{fig:maps_gaussian_gran_sasso} and \ref{fig:maps_lmc_gran_sasso}. The left panel of the figure shows the case of $m_\chi=5$~GeV and $E_R=3$~keV, while the right panel shows the case of  $m_\chi=100$~GeV and $E_R=0.25$~keV. We use a different color palette in this figure to highlight the differences, since $\epsilon(\hat{\bf q})$ spans from small values of $\sim 10^{-5}$ up to $\sim 1$. The minimum of the palette is adjusted to zero in order to provide a more gradual color transition. The yellow centered dot  indicates the direction at which the recoil rate computed assuming the local DM velocity distribution of the MW-LMC analogue is maximum. The circles around this maximum rate direction denote angular regions of 5$^\circ$ (white), 10$^\circ$ (gray), and 30$^\circ$ (black). These correspond to the angular resolutions required by the CYGNUS-like experiment ($\leq$ 30$^\circ$) to achieve event‑level reconstruction of nuclear recoil directions \cite{Vahsen:2021gnb, Lisotti:2024fco}, and represent idealized angular resolution scenarios.

In both panels of figure~\ref{fig:rel_diff}, the maximum of the relative difference, $\epsilon(\hat{\bf q})$, occurs very close to the maximum of the differential recoil rate computed assuming the DM velocity distribution from the MW-LMC analogue. Comparing the left panels of figures~\ref{fig:maps_gaussian_gran_sasso} and \ref{fig:maps_lmc_gran_sasso}, we can see that the main difference between the recoil maps is in and around the direction of $-{\bf V}_{\rm lab}$, which results in the relative differences seen in the left panel of figure~\ref{fig:rel_diff} also maximizing in that direction. We also note a considerable difference in regions where fewer events are expected. This can be understood by observing that increasing the number of DM particles with a preferred velocity direction, as in the case of the MW-LMC analogue, reduces the fraction of particles scattering in directions nearly opposite to $-\mathbf{V}_{\rm{lab}}$.

Comparing the right panels of figures~\ref{fig:maps_gaussian_gran_sasso} and \ref{fig:maps_lmc_gran_sasso}, we can see that the numerical value of the maximum rate computed assuming the DM velocity distribution from the simulated MW-LMC analogue is more than twice the maximum rate computed for the SHM. This leads to large relative differences between the maps around the directions where the rate assuming the DM velocity distribution of the MW-LMC analogue is large, as seen in the right panel of figure~\ref{fig:rel_diff}.

To quantify the relative difference, $\epsilon({\hat{\bf q}})$, which can be resolved given the angular resolution of future directional detectors, we produce directional recoil maps assuming a DM mass of $m_\chi=100$~GeV and a recoil energy ranging from 0.25 to 50~keV for both the SHM and the MW-LMC analogue velocity distribution. We then evaluate the relative difference in the two maps using eq.~\eqref{eq:rel_diff}. We identify the direction where the differential recoil rate for the MW-LMC analogue is maximum, and find the average relative difference, $\bar\epsilon$, over angular regions of $5^\circ$, $10^\circ$, and $30^\circ$ from this maximum rate direction. Notice that we consider recoil energies beyond the 10 keV upper limit considered in CYGNUS to study how the average relative difference changes for large values of $v_{\rm min}$.

\begin{figure}[t]
    \centering
    \includegraphics[width=0.7\textwidth]{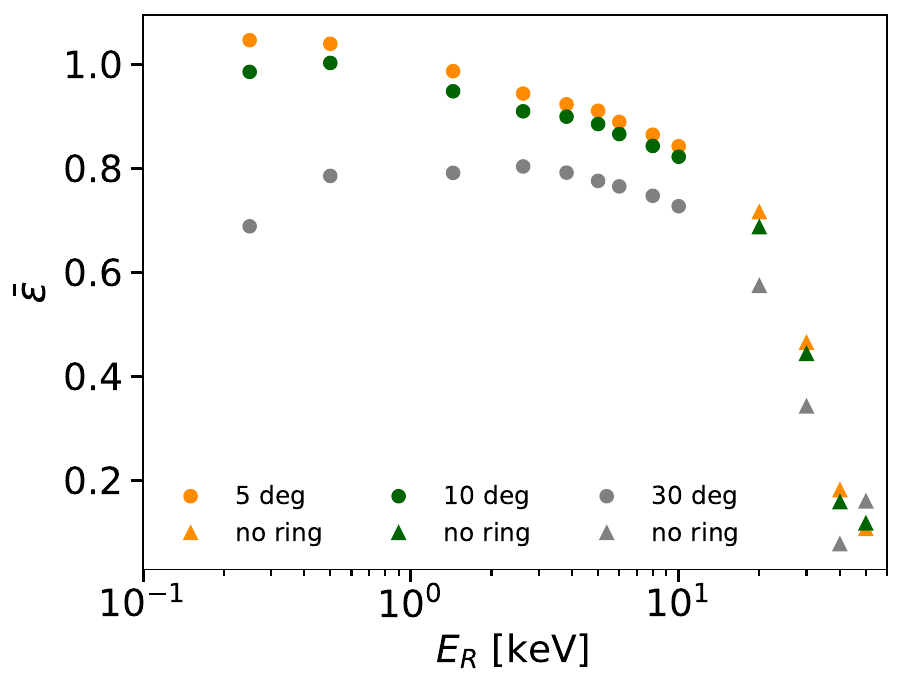}
    \caption{Average relative difference, $\bar{\epsilon}$, of the directional recoil rates computed assuming the SHM and the velocity distribution from the simulated MW-LMC analogue,  averaged over a region of $5^\circ$ (orange), $10^\circ$ (green) and $30^\circ$ (grey) from the direction where the rate computed for the MW-LMC analogue is maximum. The DM mass is assumed to be $m_\chi=100$~GeV and  $\bar{\epsilon}$ is shown for different recoil energies, $E_R$. The circles and triangles correspond to cases in which a ring does and does not appear in the SHM recoil map, respectively.}
    \label{fig:Res_vs_er}
\end{figure}

Figure~\ref{fig:Res_vs_er} shows the average relative difference, $\bar\epsilon$, as a function of recoil energy for the three angular regions of $5^\circ$, $10^\circ$, and $30^\circ$, shown as orange, green, and gray circles and triangles, respectively. Circles correspond to cases in which a ring-like feature is present in the recoil map of the SHM distribution, while triangles (noted as ``no ring" in the figure) correspond to cases where the maximum of the recoil rate is in the direction $-\mathbf{V}_{\rm{lab}}$.

As shown in figure~\ref{fig:Res_vs_er}, as the angular resolution of the experiment decreases (i.e.~as the size of the angular region that we average the relative difference over increases), the average relative difference between the rates computed for the SHM and the distribution from the MW-LMC analogue decreases. In other words, the experiment becomes less capable of resolving the region around the maximum of the directional recoil rate, as expected. Nevertheless, even with an angular resolution of $30^\circ$, relative differences can reach $\sim 80\%$ for recoil energies less than 10~keV, where the ring exists in the SHM recoil map. For larger recoil energies, the maximum of the recoil rate occurs in the direction of $-{\bf V}_{\rm lab}$ for both the SHM and the MW-LMC analogue, and the average relative difference between the two decreases.

These results indicate that the impact of the LMC on the local DM velocity distribution can produce sizeable and potentially observable modifications in the directional recoil signal, especially for $v_{\rm min}<|{\bf V}_{\rm lab}|$. Even when the detector angular resolution is limited, the deviations from the SHM remain significant over a broad range of recoil energies, suggesting that future directional experiments could, in principle, be sensitive to such non-standard halo features.

\subsection{Statistical analysis and isotropy rejection}
\label{sec:statistical_analysis}

We develop a Monte Carlo (MC) likelihood framework to estimate the number of  events required to  distinguish a DM-induced directional signal from an isotropic distribution using HEALPix  maps. The analysis is based on a likelihood-ratio test applied to simulated event samples.

Directional recoil maps are discretized in HEALPix format and we interpret them as probability distributions on the sphere. Each map is normalized as
\[
p_i = \frac{m_i}{\sum_j m_j}\,,
\]
where $p_i$ is the probability at pixel $i$, \(m_i\) is the value of the directional differential recoil rate at pixel \(i\), and the sum in the denominator is over all pixels.

For a given probability distribution \(p\), event samples of size $N$ are generated by drawing pixel indices according to the multinomial probability distribution defined by \(p\). These sampled indices constitute a mock dataset \(\{x_k\}\) of recoil events. The log-likelihood under a model \(p\) is then
\[
\log \mathcal{L}(p)
=
\sum_{k=1}^{N} \log p(x_k)\, .
\]

To test for the presence of a signal component, we define a mixture model including the isotropic background, \(p_{\rm null}\), and the signal map, \(p_{\rm sig}\),
\[
p(\mu)
=
(1-\mu)\,p_{\rm null}
+
\mu\, p_{\rm sig}\, ,
\]
where \(\mu \in [0,1]\) is the signal fraction. For each  dataset, the likelihood is evaluated over a grid of \(\mu\) values uniformly spanning the interval \([0,1]\).

We define the test statistic 
\[
{\rm TS}=2\left[\max_{\mu}\log \mathcal{L}(p(\mu))- \log \mathcal{L}(p_{\rm null})\right]\,,
\]
which quantifies how strongly the data favor the signal-plus-background hypothesis relative to the pure background hypothesis.

For each event count $N$, we generate 1000 MC realizations by drawing events from the mixture distribution $p(\mu)$. For each realization, the TS is computed by maximizing the log-likelihood over a grid of $\mu$ values uniformly spanning $[0,1]$. The isotropic component $p_{\rm null}$ enters the test statistic both as the $\mu = 0$ limit of the mixture model and as the reference log-likelihood $\log\mathcal{L}(p_{\rm null})$. We take the median TS over all realizations, ${\rm TS}_{\rm med}$, and define the expected significance $Z = \sqrt{{\rm TS}_{\rm med}}$. The minimum number of events required to reject isotropy is the smallest $N$ for which $Z \geq 3$, corresponding to a $3\sigma$ significance. More generally, we compute $Z$ as a function of $N$ and quote results for different significance thresholds.

\begin{figure}[t]
    \centering
    \begin{subfigure}{0.47\textwidth}
        \centering
        \includegraphics[width=\linewidth]{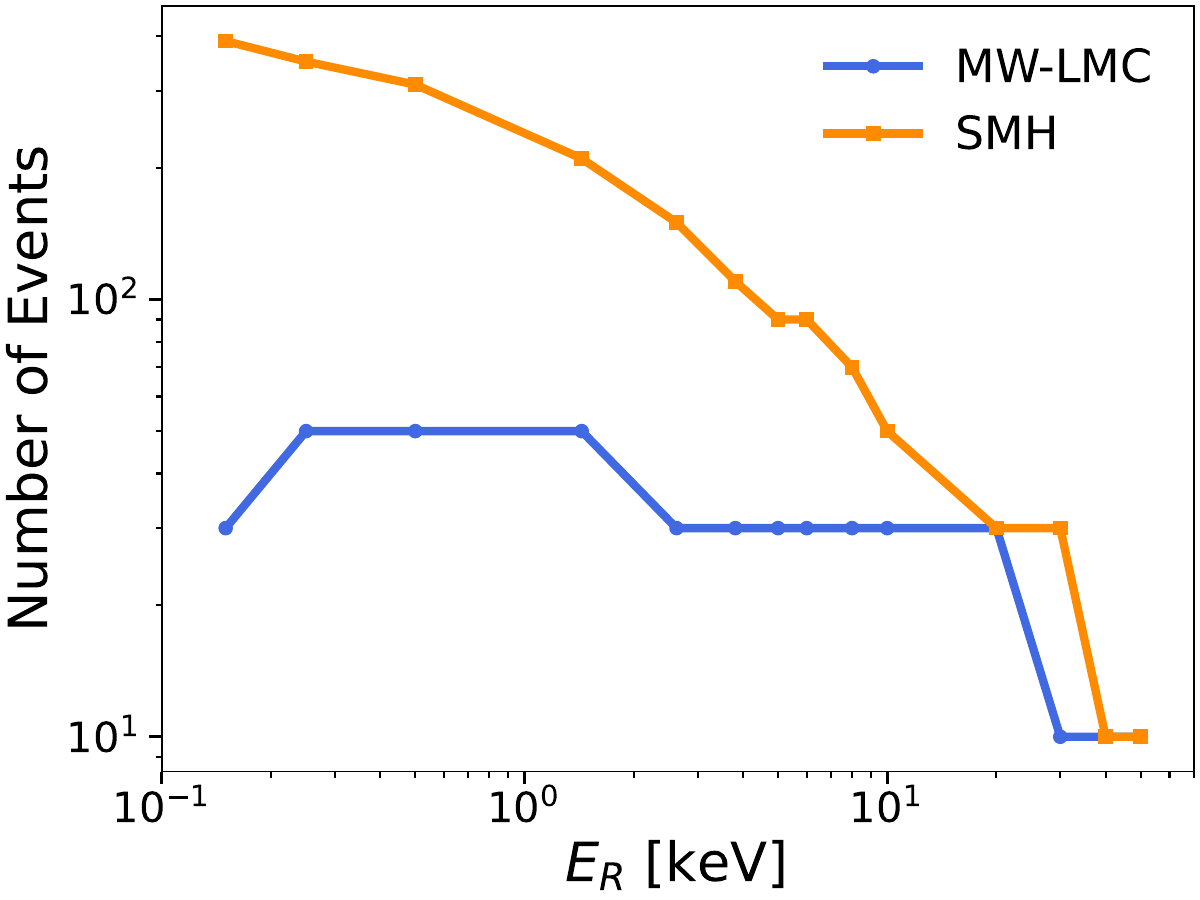}
    \end{subfigure}
     \hspace{0.4em} 
    \begin{subfigure}{0.47\textwidth}
        \centering
        \includegraphics[width=\linewidth]{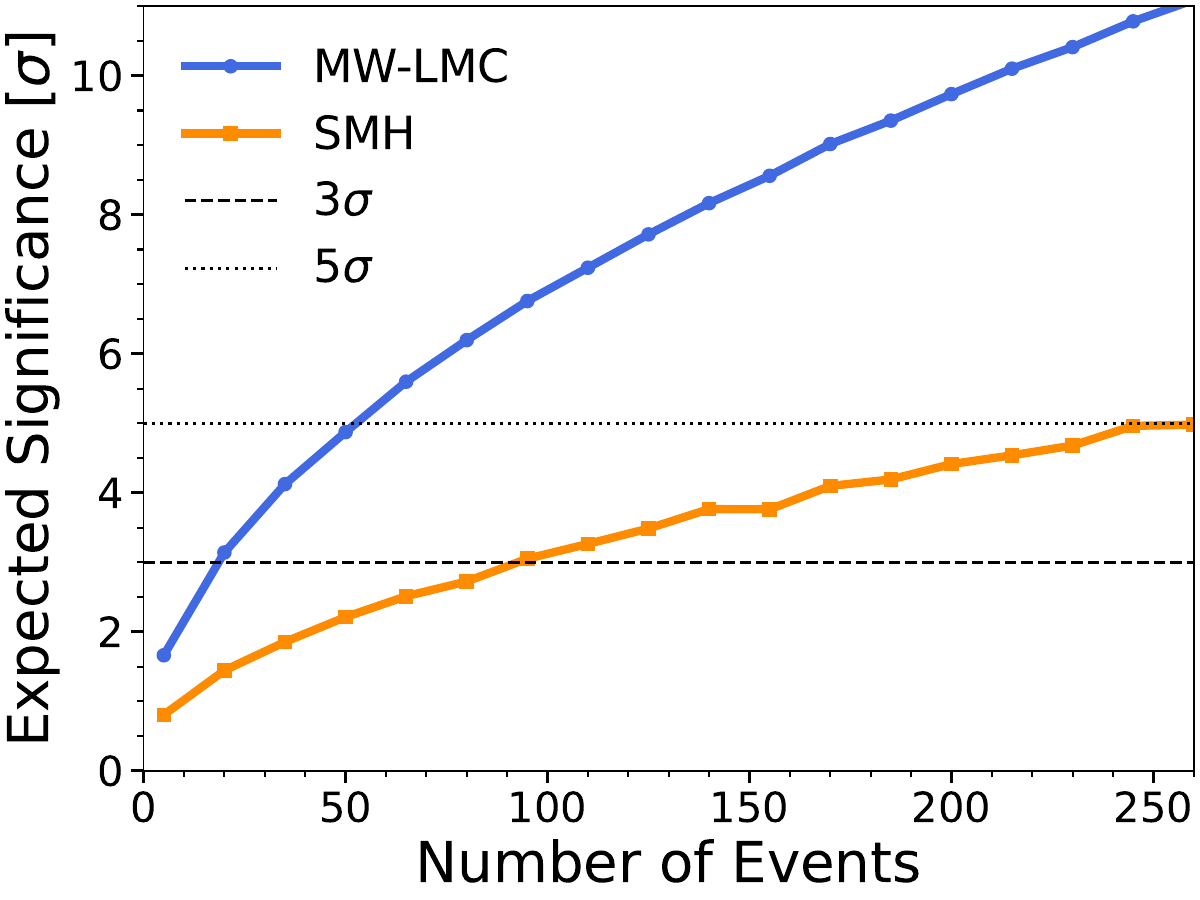}
    \end{subfigure}
    \caption{Left panel: Number of events required to reject isotropy at $3\sigma$ considering the directional recoil map
   at specific recoil energies, $E_{R}$. Right panel: Expected statistical significance as a function of the  number of events in the recoil energy interval \([0.25-10]~\mathrm{keV}\) required to reject isotropy. In both panels we adopt a DM mass of $m_\chi=100$~GeV. The orange and blue curves show the results for the SHM and the simulated MW-LMC analogue, respectively, while the dashed and dotted horizontal lines in the right panel indicate the \(3\sigma\) and \(5\sigma\) significance levels, respectively.}    
    \label{fig:events_TS}
\end{figure}

The results of our likelihood-ratio test are shown in figure~\ref{fig:events_TS}. First we perform the analysis for directional recoil maps at fixed recoil energies. The left panel of  figure~\ref{fig:events_TS} shows the number of events required to reject isotropy at  $3\sigma$ considering the directional recoil map computed assuming the SHM DM velocity distribution (orange) and the DM velocity distribution extracted from the simulated MW-LMC analogue (blue) for a DM mass of $m_\chi=100$~GeV and specific recoil energies, $E_R$. 

Next, we perform an energy-integrated analysis by stacking the directional recoil maps over the recoil energy range \([0.25-10]~\mathrm{keV}\). The energy-integrated maps are constructed pixel-by-pixel using trapezoidal integration across maps at different recoil energies, and then normalizing the resulting integrated maps. The right panel of figure~\ref{fig:events_TS} shows the expected statistical significance as a function of the required number of events in the recoil energy interval \([0.25-10]~\mathrm{keV}\) to reject isotropy. The orange and blue curves show the results  for the SHM and the simulated MW-LMC analogue, respectively.

The results of this analysis reveal a significant improvement in the detectability of the directional DM signal when the impact of the LMC is included. As shown in the left panel of figure~\ref{fig:events_TS}, rejecting isotropy requires significantly less events for the MW-LMC analogue compared to the SHM, across recoil energies less than $\sim 20$~keV. For lower recoil energies, the ring-like feature is present in the recoil maps for the SHM. However, the local DM  distribution of the MW-LMC analogue produces  distinct directional features in the recoil map (as seen in the right panel of figure~\ref{fig:maps_lmc_gran_sasso}) that are easier to  detect compared to the isotropic ring. This trend is further confirmed by the energy-integrated analysis shown in the right panel of figure~\ref{fig:events_TS}. For the MW-LMC analogue, isotropy can be rejected at the $3\sigma$ and $5\sigma$ levels with $\sim 19$ and $\sim 53$ events, respectively, while the SHM  requires  $\sim 93$ and $\sim 261$ events to reject isotropy at the same significance levels. 

\begin{figure}[t]
    \centering
    \includegraphics[width=0.7\textwidth]{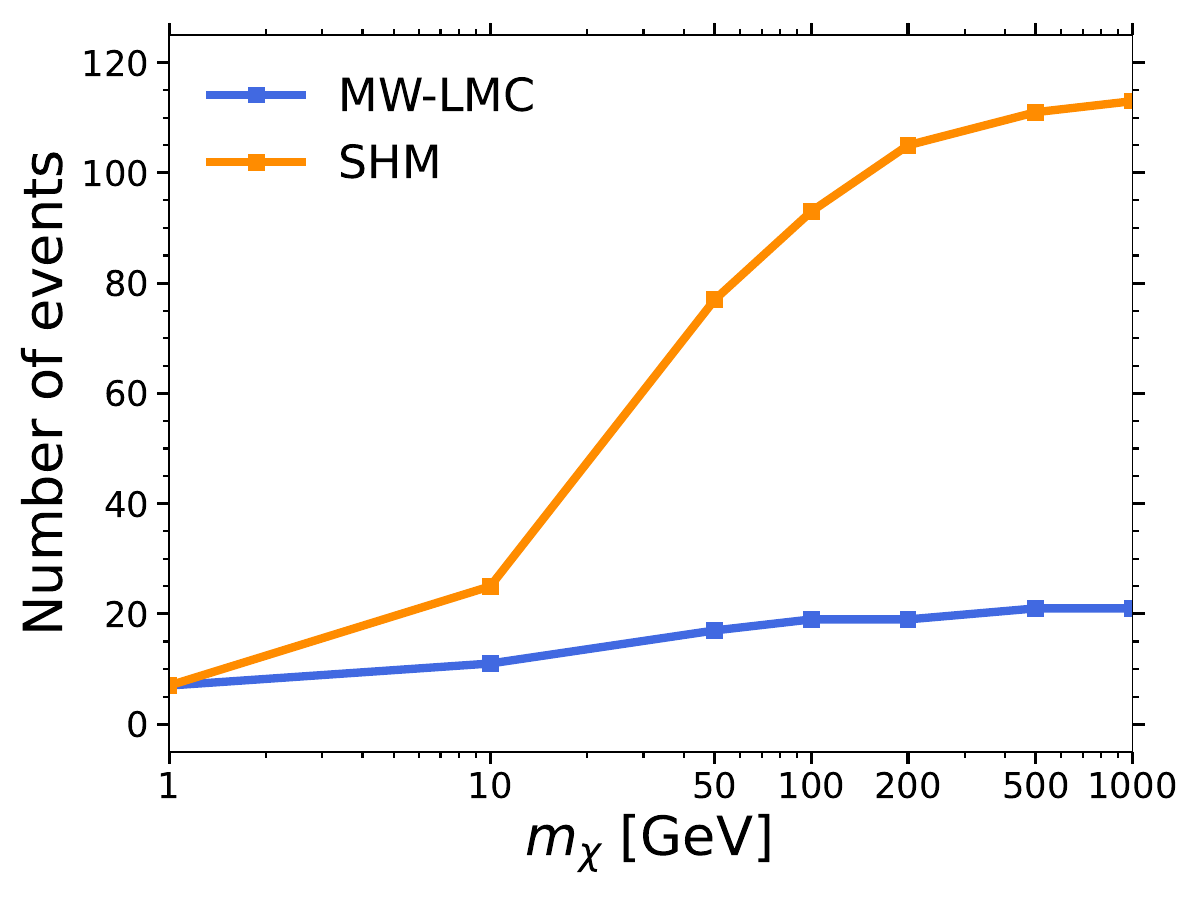}
    \caption{Number of events required to reject isotropy at $3\sigma$ as a function of the DM mass $m_\chi$, for the SHM (orange) and the MW-LMC analogue (blue), using energy-integrated directional recoil maps over $E_R=[0.25-10]~\mathrm{keV}$.}
    \label{fig:N_vs_mass}
\end{figure}

Finally, we extend the energy-integrated analysis across different values of the DM mass. Figure~\ref{fig:N_vs_mass} shows the number of events required to reject isotropy at $3\sigma$ as a function of $m_{\chi}$, for both the SHM (orange) and the MW-LMC analogue (blue) using the energy-integrated directional recoil maps over the recoil energy range $E_R=[0.25-10]$~keV. In both cases, the number of required events increases with $m_{\chi}$, which can be understood from the kinematics of the scattering. Lighter DM particles require a higher minimum velocity, $v_{\rm{min}}$, to produce a detectable recoil, and consequently $v_{\rm min}>|{\bf V}_{\rm lab}|$. This leads to the maximum recoil rate to occur in the direction of $-{\bf V}_{\rm lab}$, creating the dipole feature, which is easier to detect. Larger DM masses ($m_\chi \gtrsim 10$~GeV) lead to smaller $v_{\rm{min}}$ and produce the the ring-like feature in the directional recoil map for the SHM, which requires more events to be detected. For these masses, the LMC distorts the ring and the resulting anisotropic feature requires fewer events to reject isotropy compared to the SHM. For $m_\chi \gtrsim 100$~GeV, the value of $v_{\rm min}$ only slightly changes and the recoil maps remain similar, leading to a plateau in the curves. 

In our analysis we adopt an  angular  resolution corresponding to a HEALPix pixel size of $\sim 3^\circ$. We have also verified that using a coarser resolution of $30^\circ$ yields a required number of events to reject isotropy at a given significance level that is in very good agreement with our results.

These results highlight that the LMC can significantly enhance the prospects for  directional DM detection, reducing the required number of events  to reject isotropy, or equivalently the required experimental exposure, by a factor of several compared to the SHM. The impact of the LMC on the directional recoil rate is strongest for large DM masses and low recoil energies, where it distorts the ring-like feature in the directional recoil map.

\section{Discussion and conclusions}
\label{sec:summary}

In this work, we have investigated how the presence of the LMC reshapes the directional DM signal expected in a future CYGNUS-like directional experiment, using a simulated
MW-LMC analogue from the Auriga magneto-hydrodynamical simulations. Using the local DM velocity distribution extracted from the simulations, we have computed the expected directional recoil rate in the CYGNUS-like experiment and compared it to the standard expectation obtained from the SHM. Our results show that the LMC leaves a clear and potentially observable imprint on the angular distribution of the recoil signal.

The local DM velocity distribution in the Galactic frame obtained from the simulated MW--LMC analogue is mildly radially anisotropic, with a velocity anisotropy parameter of $\beta = 0.136$. While this indicates a departure from isotropy, our tests suggest that this radial anisotropy alone is not the primary driver of the features observed in the directional recoil maps. Instead, the dominant effect appears to arise from the azimuthal component of the local velocity distribution. In particular, the azimuthal velocity component peaks at
$u_\phi \sim  75$~km/s, indicating a counter-rotating DM component relative to the stellar disk, due to the impact of the LMC. 

The impact of the LMC on the directional recoil signal depends strongly on the minimum DM speed, $v_{\rm min}$, as compared to the laboratory speed, $|{\bf V}_{\rm lab}|$. For $v_{\rm min} > |{\bf V}_{\rm lab}|$, the direction of the maximum recoil rate remains aligned with $-{\bf V}_{\rm lab}$, as in the SHM. Consequently, the relative differences between the recoil maps obtained from the SHM and the MW-LMC analogue are concentrated around the direction $-{\bf V}_{\rm lab}$. By contrast, when $v_{\rm min} < |{\bf V}_{\rm lab}|$, the effect of the LMC is qualitatively different. In the SHM, this regime produces a ring-like feature, where the maximum recoil rate occurs in a ring around the direction $-{\bf V}_{\rm lab}$. We find that the inclusion of the LMC distorts this ring entirely. Rather than preserving azimuthal symmetry around $-{\bf V}_{\rm lab}$, the recoil rate becomes concentrated at preferred azimuthal angles, generating a distinct asymmetric pattern. 

To further quantify these effects, we have computed the relative difference between the recoil maps for the MW-LMC analogue and the SHM. We find that such a relative difference reaches values as large as $\sim 80\%$ in regions in the directional recoil map close to the maximum of the rate, even after averaging over angular regions of $30^\circ$ as seen in figure~\ref{fig:rel_diff}. For a DM mass of 100 GeV, the largest deviations occur at recoil energies $\lesssim 10~\mathrm{keV}$, where the ring-like feature is present in the SHM recoil map. This indicates that the LMC-induced effects could potentially be detectable 
with future directional experiments even for lower angular resolutions.

We have also quantified the impact of the LMC on the discovery reach of the CYGNUS-like experiment through a likelihood-based isotropy rejection analysis. We find that including the LMC substantially improves the ability of the experiment to distinguish a DM signal from an isotropic background. For a DM mass of $m_\chi=100$~GeV and for the energy-integrated analysis over the range of $E_R =[0.25-10]~\mathrm{keV}$, isotropy can be rejected at $3\sigma$ and $5\sigma$ significance with approximately $\sim 19$ and $\sim 53$ events, respectively, for the MW-LMC analogue. In comparison, the SHM requires approximately $\sim 93$ and $\sim 261$ events to reach the same significance levels, respectively. The LMC therefore reduces the required number of events to reject isotropy by nearly a factor of 5 for $m_\chi=100$~GeV. This reduction becomes even more pronounced at larger DM masses, exceeding a factor of 5. 
This translates directly into a lower required experimental exposure, with the largest gains occurring for high DM masses and low recoil energies, where the ring-like feature is present in the SHM but distorted due to the impact of the LMC. 

The analysis presented here is based on a single MW--LMC analogue from the Auriga cosmological simulations. Extending this study to a larger sample of simulated halos with different LMC orbital histories and masses would help quantify the halo-to-halo variation of these effects. Nevertheless, the strong directional signatures found in this work indicate that the LMC should be considered an important astrophysical ingredient in future predictions for directional DM detection.

\acknowledgments We thank Azadeh Fattahi for providing the re-simulated Auriga halo studied in this work. JRC and NB acknowledge the support of the Natural Sciences and Engineering Research Council of Canada (NSERC), funding reference number RGPIN-2020-07138, and the NSERC Discovery Launch Supplement, DGECR-2020-00231. JRC acknowledges support from INFN through the Senior Research Fellowship 
program (Grant No.~27076). NB acknowledges the support of the Canada Research Chairs Program.

\appendix
\section{Tests on the numerical integration}
\label{sec:appendix_numerical_int}

In this appendix, we present the results of the tests performed to determine the accuracy of our numerical approach for integrating the DM velocity distribution obtained from the MW-LMC analogue.

\textbf{Maxwellian distribution}. We compare the results of our numerical framework with the analytical expression of the Radon transform for the Maxwellian distribution, given in eq.~\eqref{eq:radon_gaussian}. For this test, we create mock numerical data by sampling $N$ independent values of the local DM velocities in the Galactic frame, $(u_x,u_y,u_z)$, from a truncated  Maxwellian velocity distribution, and then apply the methodology detailed in section~\ref{sec:numerical_radon}. We assume a DM mass of 100 GeV and a nuclear recoil energy of 0.25 keV as our benchmark case, with the specifications of the CYGNUS-like experiment at Gran Sasso for a He target. 

\begin{figure}[t]
    \centering
    \begin{subfigure}{0.47\textwidth}
        \centering
        \includegraphics[width=\linewidth]{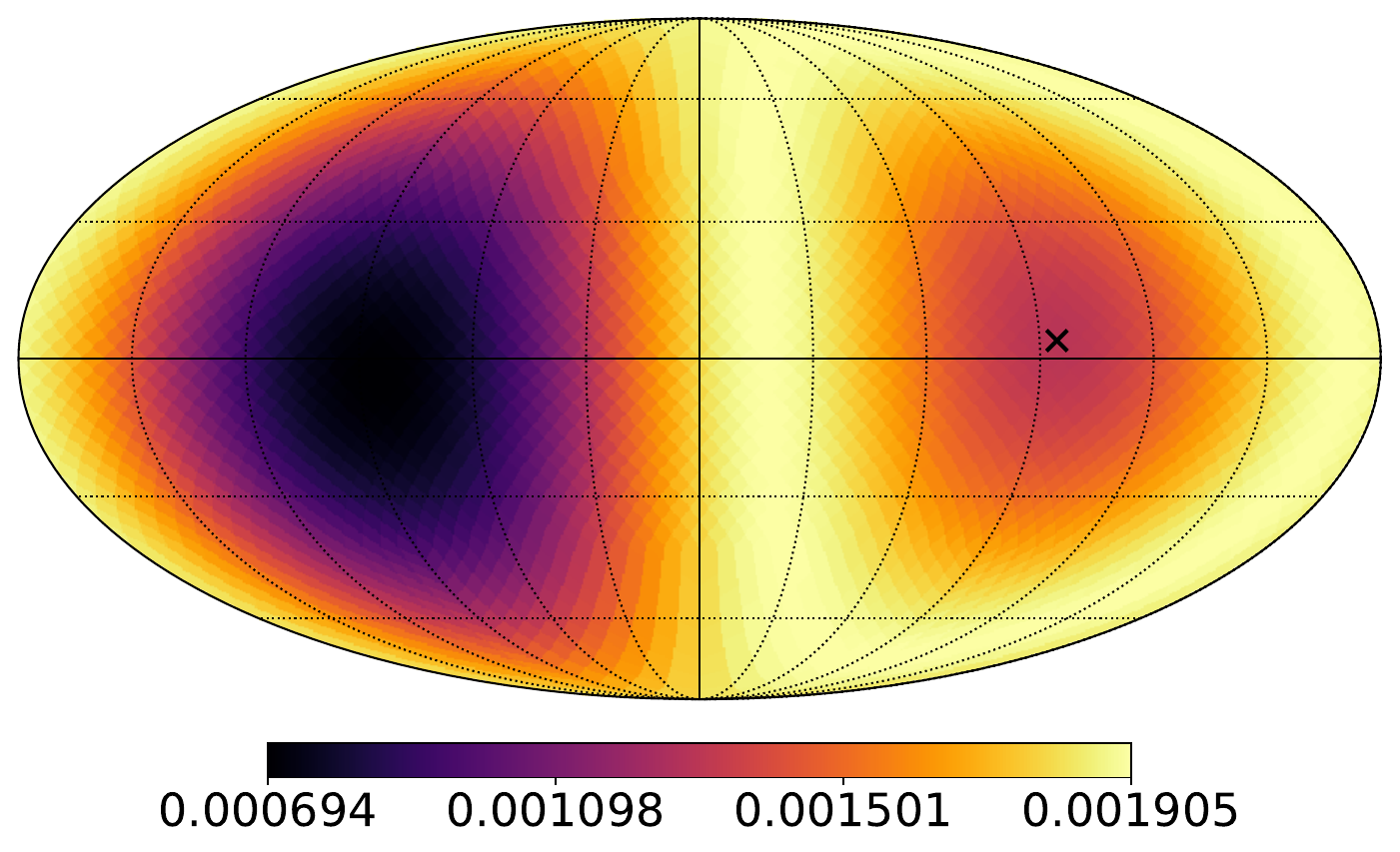}
    \end{subfigure}
     \hspace{0.4em} 
    \begin{subfigure}{0.47\textwidth}
        \centering
        \includegraphics[width=\linewidth]{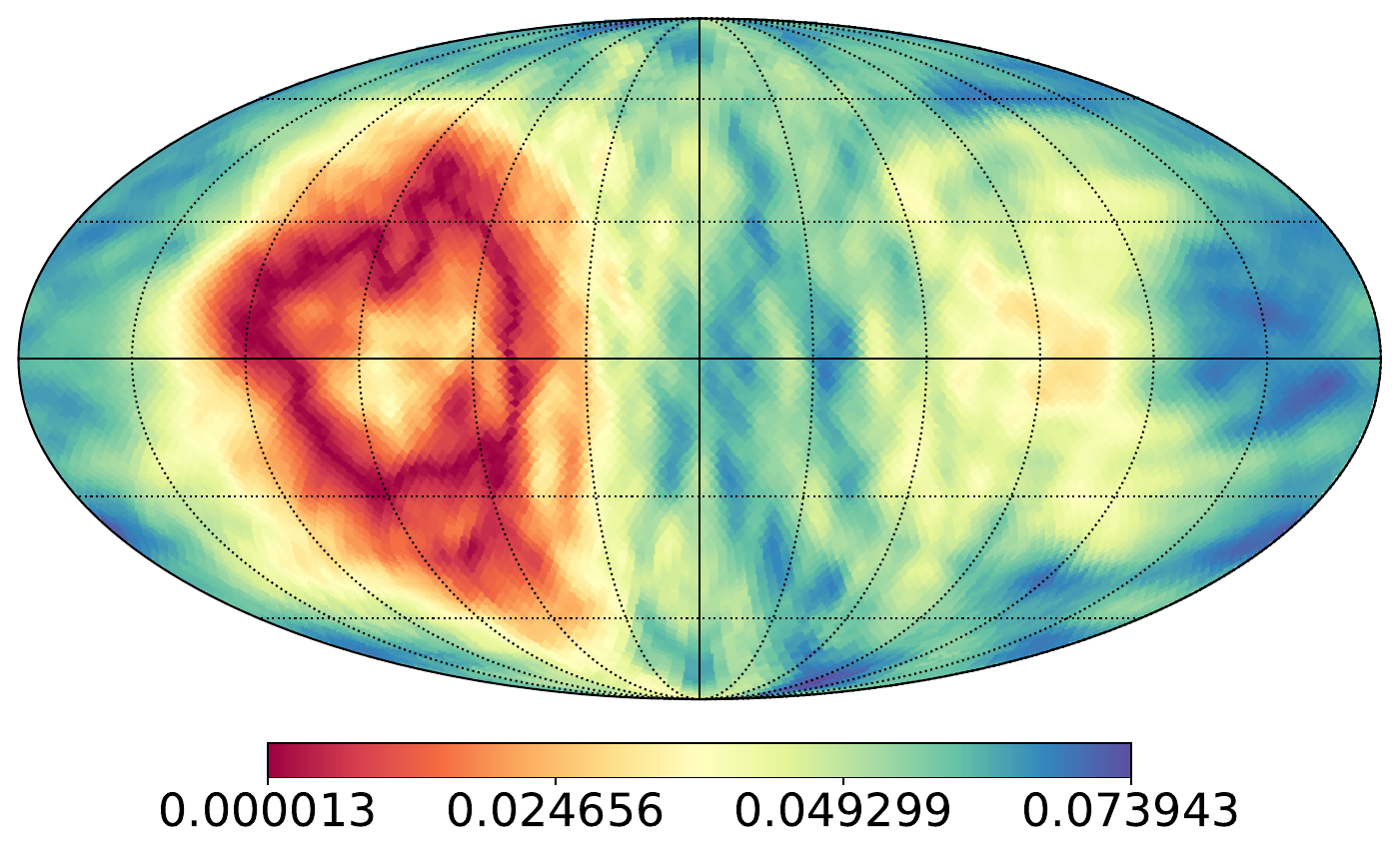}
    \end{subfigure}
    \caption{Left panel: Molleweide map of $\hat f_{\rm det}$ in Galactic coordinates for a He target at Gran Sasso assuming the SHM Maxwellian DM velocity distribution and computed using the exact expression in eq.~\eqref{eq:radon_gaussian},  for $m_\chi=100$ GeV and $E_R=0.25$~keV. Right panel: Relative difference of the $\hat f_{\rm det}$ maps computed using the numerical integration from mock data  created from a truncated Maxwellian distribution and the exact expression of eq.~\eqref{eq:radon_gaussian}.}
    \label{fig:test_num_integration}
\end{figure}

The Mollweide map of $\hat{f}_{{\rm det}}(v_{\rm min},\hat{\mathbf{q}})$ computed using eq.~\eqref{eq:radon_gaussian}  is shown in the left panel of figure~\ref{fig:test_num_integration}. 
We compare this exact result, $\hat f_{\rm exact}$, with the result of our numerical integration, $\hat f_{\rm num}$, by computing their relative difference defined as  $\epsilon(\hat{\bf q}) = \left|1 - \hat f_{\rm num}/\hat f_{\rm exact}\right|$. We show this relative difference in the right panel of figure~\ref{fig:test_num_integration}. To create the $\hat f_{\rm num}$ map, we use a total of $10^4$ DM particles, since this is approximately the size of the simulation sample. As it can be seen from the right panel of figure~\ref{fig:test_num_integration}, the relative difference between the numerical and exact $\hat f_{\rm det}$ maps is at most $\sim 7.4$\%.

Within the same test, we also vary several relevant parameters of the numerical framework:

\begin{itemize}

    \item \textbf{Number of particles}. The sample size of the DM velocity distribution can play an important role in the numerical framework, as a larger number of sampled particles may provide a more accurate reconstruction of the distribution. As mentioned previously, we compute $\hat{f}_{\rm{det}}(v_{\rm min},\hat{\mathbf{q}})$ from a numerical sample in which each $(u_x,u_y,u_z)$ component is drawn from a Maxwellian distribution, using $ 10^4$ particles to match the approximate number of particles in the simulation. We then increase the number of sampled particles up to $\sim 10^6$. We find that this produces a maximum relative difference of $\sim 7.1$\% in $\hat f_{\rm det}$, with respect  to the exact solution, 
   which is very close to the maximum value shown in the right panel of figure~\ref{fig:test_num_integration}. We therefore conclude that, within this range, the sample size does not significantly affect our results.

    \item \textbf{Bin Size}. The size of the velocity bins can also affect the reconstructed numerical velocity distribution obtained from the sampled particles. As described in section~\ref{sec:numerical_radon}, we adopt a bin size of $\Delta v = 50~\rm{km}/\rm{s}$ for our benchmark computations. We then vary the bin size and compare the resulting distributions, choosing $\Delta v = 25$~km/s and $\Delta v = 75$~km/s. For $N=10^{4}$ sampled particles, a bin size of $\Delta v = 25$ km/s yields results very similar to the benchmark case; however, the distribution at small and large velocities is less well-represented because the number of particles per bin decreases. For $\Delta v = 75$~km/s, the numerical computation shows a larger mismatch with respect to the exact solution, with the relative difference reaching a maximum of $\sim 8.3 \%$. We conclude that larger velocity bins reduce the resolution of the reconstructed distribution, leading to a poorer representation of the underlying velocity structure.

    \item \textbf{Integrations points}. The computation of the Radon transform in eq.~\eqref{eq:integral_to_perform} requires evaluating a double numerical integral for each pixel in the Mollweide map, which can be computationally expensive. For example, when the angular resolution of the HEALPix grid is $\sim 2^\circ$, corresponding to  ${\rm Nside} = 32$, the total number of pixels is 12,288. To keep the computational cost manageable, we use a trapezoidal integration scheme. The total runtime depends strongly on the number of integration points used. We find that using 100 integration points is sufficient to ensure convergence, as increasing this number changes the results by no more than $\mathcal{O}(10^{-3}\%)$.

    \item \textbf{Nside}. We have also verified that the HEALPix angular resolution  does not significantly affect the resulting maps. In particular, we achieve the desired accuracy already at ${\rm Nside} = 16$, which further reduces the computational time. Nevertheless, we adopt ${\rm Nside} = 32$ in the final results for aesthetic purposes, as it provides a smoother visual representation without altering the physical conclusions.
\end{itemize}

\section{Resolution tests}
\label{sec:lmc_sampling}

We have argued in appendix~\ref{sec:appendix_numerical_int} that increasing the number of sampled DM velocities in the case of a Maxwellian distribution does not have a significant impact on our results. However, this conclusion does not necessarily extend to the DM velocity distribution extracted from the simulated MW-LMC analogue. In this case, the simulation resolution imposes an intrinsic limitation, preventing us from assessing whether increasing the number of particles would significantly affect our results. A higher resolution simulation would be required to fully test the stability of our conclusions with respect to the sampling of the underlying DM velocity distribution.

Nevertheless, using MC techniques, we can artificially increase the number of points drawn from the DM velocity distribution obtained from simulations. We obtain the one-dimensional DM velocity distributions in $u_x$, $u_y$, and $u_z$, as well as the distribution of the speed $u$ in the Galactic frame, directly from the simulations, by binning the velocity data and computing the mean value and $1\sigma$ Poisson uncertainty in each bin as shown in figure~\ref{fig:velocity_components}. We assume that the measured value of each velocity component and of the total speed within a given bin follows a Gaussian distribution, with the mean given by the measured mean value in each bin and the standard deviation given by the corresponding $1\sigma$ Poisson error in that bin. Under this assumption, we construct a likelihood of the form
\begin{equation}
\log\mathcal{L} = -\frac{1}{2}\sum_{i=1}^{3} \left( \frac{u_{i} - \bar{u_{i}}}{\sigma_{i}} \right)^{2} - \frac{1}{2}\left( \frac{u - \bar{u}}{\sigma}\right)^2,
\label{eq:like_lmc}
\end{equation}
where $u_i \in [-600,600]\ \mathrm{km/s}$ with $i=x,y,z$, denotes a velocity value within the bin centered at $\bar{u}_i$, and $\sigma_i$ and $\sigma$ are the corresponding uncertainties for the individual velocity components and for the speed $u=|\mathbf{u}|$, respectively.

We sample the distribution by generating random values for the individual velocities $u_i$ and for the speed $u$, identifying the corresponding velocity bins, and computing the likelihood for those values. We use the Python package \texttt{emcee} to perform the Markov Chain MC (MCMC)  sampling. By increasing the number of accepted steps, or, in the case of \texttt{emcee}, the number of independent walkers, we effectively increase the number of particles in the reconstructed velocity distributions.

\begin{figure}[t]
    \centering
    \begin{subfigure}{0.47\textwidth}
        \centering
        \includegraphics[width=\linewidth]{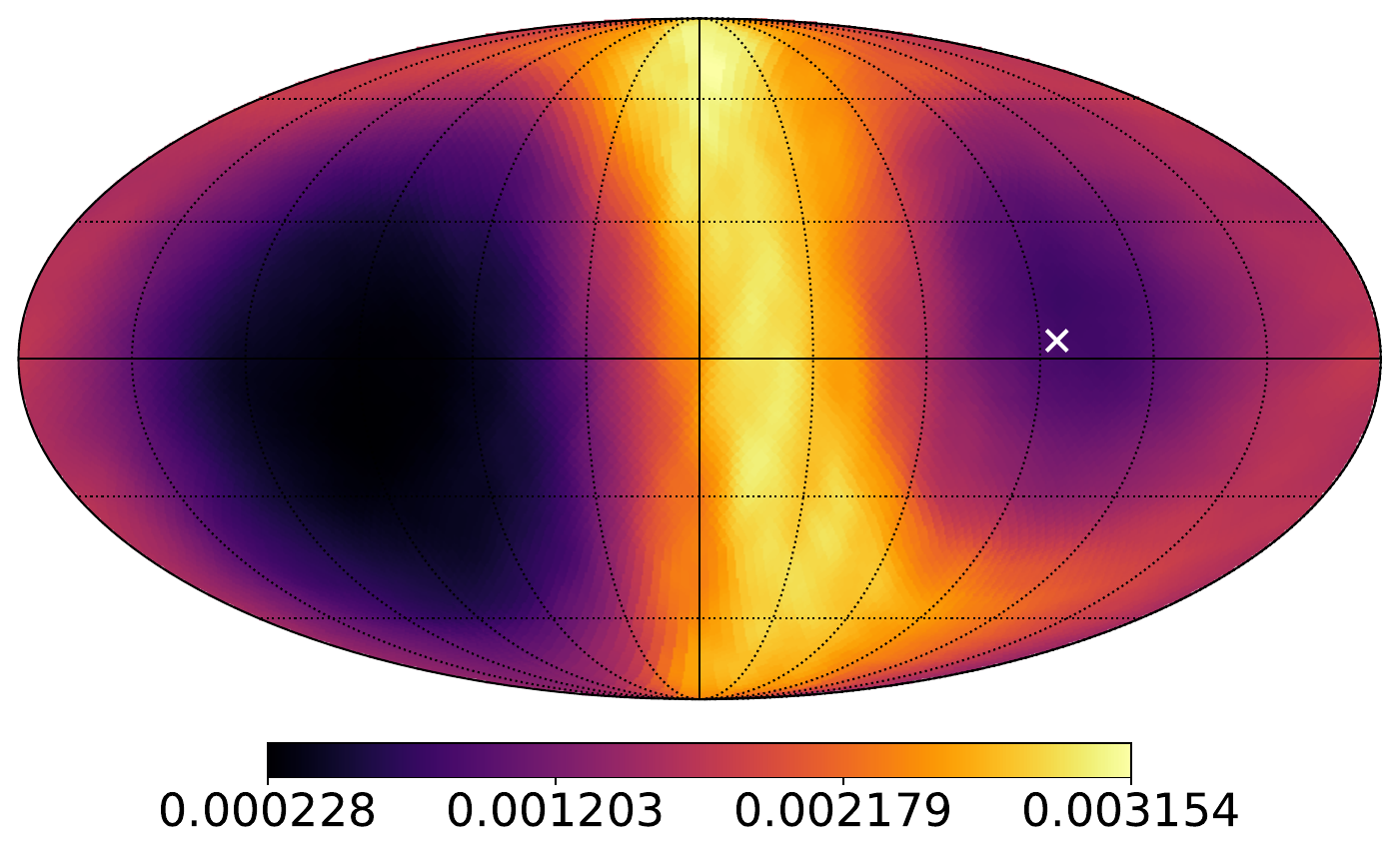}
    \end{subfigure}
     \hspace{0.4em} 
    \begin{subfigure}{0.47\textwidth}
        \centering
        \includegraphics[width=\linewidth]{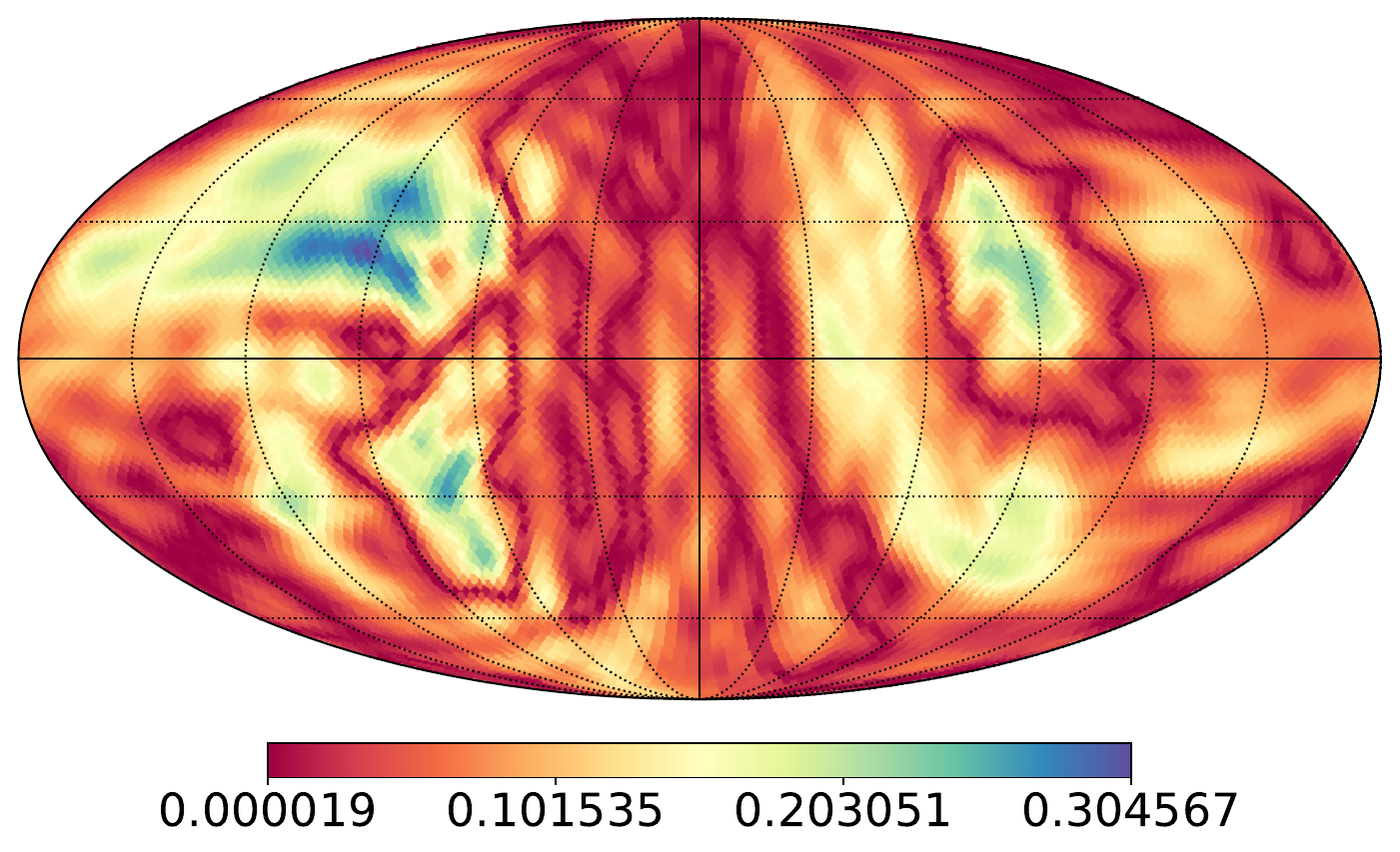}
    \end{subfigure}
    \caption{Left panel: Mollweide map of $\hat f_{\rm det}$ in Galactic coordinates for a He target at Gran Sasso, computed using the MCMC sampling. Right panel: Relative difference of $\hat f_{\rm det}$ between the original distribution obtained from the simulations and the sampled distribution.}
    \label{fig:test_lmc_sampling}
\end{figure}

In the left panel of figure~\ref{fig:test_lmc_sampling}, we show the Mollweide map of $\hat f_{\rm det}$ obtained using the sampled distribution, for a DM mass of 100 GeV, nuclear recoil energy of 0.25 keV, and a He target at the Gran Sasso laboratory. To further quantify the difference between the sampled distribution and the original one computed directly from the simulations data, we compute the map of their relative difference; the results are shown in the right panel of figure~\ref{fig:test_lmc_sampling}. We observe that the  relative difference can reach values of up to $\sim 30\%$, primarily in regions where the value of $\hat f_{\rm det}$ is very low. This relatively large difference can be understood by noting that the number of particles with the specific velocities required to produce recoils in those directions is small, and therefore the MCMC sampling cannot accurately reconstruct this low-density regime. On the other hand, the relative difference is at most $\sim 15\%$ in regions where $\hat f_{\rm det}$ reaches its maximum value.

An additional source of uncertainty introduced by this procedure arises from the assumption that the individual velocity components $(u_x, u_y, u_z)$ are independent, as implied by the likelihood construction in eq.~\eqref{eq:like_lmc}. In reality, the simulation provides the full phase space information $(x, y, z, u_x, u_y, u_z)$ of the DM particles, and correlations between velocity components are present. Moreover, the components of the velocity distribution as well as the speed distribution obtained from the simulations are not exactly Gaussian or Maxwellian, and this introduces another source of uncertainty in the procedure.

Nevertheless, we argue that even without explicitly accounting for these correlations, our results, particularly in the regions where the Radon transform is largest, are not expected to change significantly if the number of particles in the numerical sample is increased, as we have found a difference of $\sim 10^{-5}$ for these regions.

\typeout{}
\bibliographystyle{JHEP}
\bibliography{refs}

@article{Gondolo:2002np,
    author = "Gondolo, Paolo",
    title = "{Recoil momentum spectrum in directional dark matter detectors}",
    eprint = "hep-ph/0209110",
    archivePrefix = "arXiv",
    reportNumber = "CWRU-P10-02, NSF-ITP-02-95",
    doi = "10.1103/PhysRevD.66.103513",
    journal = "Phys. Rev. D",
    volume = "66",
    pages = "103513",
    year = "2002"
}

@article{Cirelli:2024ssz,
    author = "Cirelli, Marco and Strumia, Alessandro and Zupan, Jure",
    title = "{Dark Matter}",
    eprint = "2406.01705",
    archivePrefix = "arXiv",
    primaryClass = "hep-ph",
    month = "6",
    year = "2024"
}

@ARTICLE{GaravitoCamargo:2021tcp,
       author = {{Garavito-Camargo}, Nicolas and {Besla}, Gurtina and {Laporte}, Chervin F.~P. and {Johnston}, Kathryn V. and {G{\'o}mez}, Facundo A. and {Watkins}, Laura L.},
        title = "{Hunting for the Dark Matter Wake Induced by the Large Magellanic Cloud}",
      journal = {Astrophys. J.},
         year = 2019,
        month = oct,
       volume = {884},
       number = {1},
          eid = {51},
        pages = {51},
          doi = {10.3847/1538-4357/ab32eb},
archivePrefix = {arXiv},
       eprint = {1902.05089},
 primaryClass = {astro-ph.GA},
       adsurl = {https://ui.adsabs.harvard.edu/abs/2019ApJ...884...51G}
}

@article{Garavito-Camargo:2020lqm,
    author = "Garavito-Camargo, Nicolas and Besla, Gurtina and Laporte, Chervin F. P. and Price-Whelan, Adrian M. and Cunningham, Emily C. and Johnston, Kathryn V. and Weinberg, Martin D. and Gomez, Facundo A.",
    title = "{Quantifying the Impact of the Large Magellanic Cloud on the Structure of the Milky Way\textquoteright{}s Dark Matter Halo Using Basis Function Expansions}",
    eprint = "2010.00816",
    archivePrefix = "arXiv",
    primaryClass = "astro-ph.GA",
    doi = "10.3847/1538-4357/ac0b44",
    journal = "Astrophys. J.",
    volume = "919",
    number = "2",
    pages = "109",
    year = "2021"
}

@article{Petersen_2020,
	doi = {10.1038/s41550-020-01254-3},
  
	url = {https://doi.org/10.1038\%2Fs41550-020-01254-3},
  
	year = 2020,
	month = {nov},
  
	publisher = {Springer Science and Business Media {LLC}
},
  
	volume = {5},
  
	number = {3},
  
	pages = {251--255},
  
	author = {Michael S. Petersen and Jorge Pe{\~{n}}arrubia},
  
	title = {Detection of the Milky Way reflex motion due to the Large Magellanic Cloud infall},
  
	journal = {Nature Astronomy}
}

@article{Bozorgnia:2016qkh,
    author = "Bozorgnia, Nassim and Gelmini, Graciela B. and Gondolo, Paolo",
    title = "{Inverted dipole feature in directional detection of exothermic dark matter}",
    eprint = "1611.01750",
    archivePrefix = "arXiv",
    primaryClass = "astro-ph.CO",
    doi = "10.1088/1475-7516/2017/01/052",
    journal = "JCAP",
    volume = "01",
    pages = "052",
    year = "2017"
}

@article{Cunningham:2020nlo,
    author = "Cunningham, Emily C. and Garavito-Camargo, Nicolas and Deason, Alis J. and Johnston, Kathryn V. and Erkal, Denis and Laporte, Chervin F. P. and Besla, Gurtina and Luger, Rodrigo and Sanderson, Robyn E.",
    title = "{Quantifying the Stellar Halo's Response to the LMC's Infall with Spherical Harmonics}",
    eprint = "2006.08621",
    archivePrefix = "arXiv",
    primaryClass = "astro-ph.GA",
    doi = "10.3847/1538-4357/ab9b88",
    journal = "Astrophys. J.",
    volume = "898",
    number = "1",
    pages = "4",
    year = "2020"
}

@article{Conroy_2021,
	doi = {10.1038/s41586-021-03385-7},
  
	url = {https://doi.org/10.1038\%2Fs41586-021-03385-7},
  
	year = 2021,
	month = {apr},
  
	publisher = {Springer Science and Business Media {LLC}
},
  
	volume = {592},
  
	number = {7855},
  
	pages = {534--536},
  
	author = {Charlie Conroy and Rohan P. Naidu and Nicol{\'{a}}s Garavito-Camargo and Gurtina Besla and Dennis Zaritsky and Ana Bonaca and Benjamin D. Johnson},
  
	title = {All-sky dynamical response of the Galactic halo to the Large~Magellanic Cloud},
  
	journal = {Nature}
}

@ARTICLE{Peterson_2020_MNRAS,
       author = {{Petersen}, Michael S. and {Pe{\~n}arrubia}, Jorge},
        title = "{Reflex motion in the Milky Way stellar halo resulting from the Large Magellanic Cloud infall}",
      journal = {Mon. Not. Roy. Astron. Soc.},
         year = 2020,
        month = may,
       volume = {494},
       number = {1},
        pages = {L11-L16},
          doi = {10.1093/mnrasl/slaa029},
archivePrefix = {arXiv},
       eprint = {2001.09142},
 primaryClass = {astro-ph.GA},
       adsurl = {https://ui.adsabs.harvard.edu/abs/2020MNRAS.494L..11P}
}

@ARTICLE{Cavieres_2025,
       author = {{Cavieres}, Manuel and {Chanam{\'e}}, Julio and {Navarrete}, Camila and {Ordenes-Brice{\~n}o}, Yasna and {Garavito-Camargo}, Nicol{\'a}s and {Besla}, Gurtina and {Hempel}, Maren and {Vivas}, A. Katherina and {G{\'o}mez}, Facundo},
        title = "{The Distant Milky Way Halo from the Southern Hemisphere: Characterization of the LMC-induced Dynamical Friction Wake}",
      journal = {\apj},
         year = 2025,
        month = apr,
       volume = {983},
       number = {1},
          eid = {83},
        pages = {83},
          doi = {10.3847/1538-4357/adbf08},
archivePrefix = {arXiv},
       eprint = {2410.00114},
 primaryClass = {astro-ph.GA},
       adsurl = {https://ui.adsabs.harvard.edu/abs/2025ApJ...983...83C}
}

@article{Bozorgnia:2012eg,
    author = "Bozorgnia, Nassim and Gelmini, Graciela B. and Gondolo, Paolo",
    title = "{Aberration features in directional dark matter detection}",
    eprint = "1205.2333",
    archivePrefix = "arXiv",
    primaryClass = "astro-ph.CO",
    doi = "10.1088/1475-7516/2012/08/011",
    journal = "JCAP",
    volume = "08",
    pages = "011",
    year = "2012"
}

@article{Green:2010zm,
    author = "Green, Anne M. and Morgan, Ben",
    title = "{The median recoil direction as a WIMP directional detection signal}",
    eprint = "1002.2717",
    archivePrefix = "arXiv",
    primaryClass = "astro-ph.CO",
    doi = "10.1103/PhysRevD.81.061301",
    journal = "Phys. Rev. D",
    volume = "81",
    pages = "061301",
    year = "2010"
}

@article{Alenazi:2007sy,
    author = "Alenazi, Moqbil S. and Gondolo, Paolo",
    title = "{Directional recoil rates for WIMP direct detection}",
    eprint = "0712.0053",
    archivePrefix = "arXiv",
    primaryClass = "astro-ph",
    doi = "10.1103/PhysRevD.77.043532",
    journal = "Phys. Rev. D",
    volume = "77",
    pages = "043532",
    year = "2008"
}

@article{Bozorgnia:2024pwk,
    author = "Bozorgnia, Nassim and Bramante, Joseph and Cline, James M. and Curtin, David and McKeen, David and Morrissey, David E. and Ritz, Adam and Viel, Simon and Vincent, Aaron C. and Zhang, Yue",
    title = "{Dark matter candidates and searches}",
    eprint = "2410.23454",
    archivePrefix = "arXiv",
    primaryClass = "hep-ph",
    doi = "10.1139/cjp-2024-0128",
    journal = "Can. J. Phys.",
    volume = "103",
    number = "8",
    pages = "671--703",
    year = "2025"
}

@article{Bozorgnia:2011vc,
    author = "Bozorgnia, Nassim and Gelmini, Graciela B. and Gondolo, Paolo",
    title = "{Ring-like features in directional dark matter detection}",
    eprint = "1111.6361",
    archivePrefix = "arXiv",
    primaryClass = "astro-ph.CO",
    doi = "10.1088/1475-7516/2012/06/037",
    journal = "JCAP",
    volume = "06",
    pages = "037",
    year = "2012"
}

@article{Goodman:1984dc,
    author = "Goodman, Mark W. and Witten, Edward",
    editor = "Srednicki, M. A.",
    title = "{Detectability of Certain Dark Matter Candidates}",
    reportNumber = "Print-85-0030 (PRINCETON)",
    doi = "10.1103/PhysRevD.31.3059",
    journal = "Phys. Rev. D",
    volume = "31",
    pages = "3059",
    year = "1985"
}

@article{Santos-Santos:2023ubx,
    author = "Santos-Santos, Isabel and Bozorgnia, Nassim and Fattahi, Azadeh and Navarro, Julio F.",
    title = "{Are there any extragalactic high speed dark matter particles in the Solar neighborhood?}",
    eprint = "2308.15388",
    archivePrefix = "arXiv",
    primaryClass = "astro-ph.GA",
    doi = "10.1088/1475-7516/2024/03/046",
    journal = "JCAP",
    volume = "03",
    pages = "046",
    year = "2024"
}

@article{NEWS-G:2022kon,
    author = "Balogh, L. and others",
    collaboration = "NEWS-G",
    title = "{The NEWS-G detector at SNOLAB}",
    eprint = "2205.15433",
    archivePrefix = "arXiv",
    primaryClass = "physics.ins-det",
    doi = "10.1088/1748-0221/18/02/T02005",
    journal = "JINST",
    volume = "18",
    number = "02",
    pages = "T02005",
    year = "2023"
}

@article{Smith-Orlik:2023kyl,
    author = "Smith-Orlik, Adam and others",
    title = "{The impact of the Large Magellanic Cloud on dark matter direct detection signals}",
    eprint = "2302.04281",
    archivePrefix = "arXiv",
    primaryClass = "astro-ph.GA",
    doi = "10.1088/1475-7516/2023/10/070",
    journal = "JCAP",
    volume = "10",
    pages = "070",
    year = "2023"
}

@article{PhysRev.104.1466,
  title = {Inelastic and Elastic Scattering of 187-Mev Electrons from Selected Even-Even Nuclei},
  author = {Helm, Richard H.},
  journal = {Phys. Rev.},
  volume = {104},
  issue = {5},
  pages = {1466--1475},
  numpages = {0},
  year = {1956},
  month = {Dec},
  publisher = {American Physical Society},
  doi = {10.1103/PhysRev.104.1466},
  url = {https://link.aps.org/doi/10.1103/PhysRev.104.1466}
}

@article{PhysRevD.33.3495,
  title = {Detecting cold dark-matter candidates},
  author = {Drukier, Andrzej K. and Freese, Katherine and Spergel, David N.},
  journal = {Phys. Rev. D},
  volume = {33},
  issue = {12},
  pages = {3495--3508},
  numpages = {0},
  year = {1986},
  month = {Jun},
  publisher = {American Physical Society},
  doi = {10.1103/PhysRevD.33.3495},
  url = {https://link.aps.org/doi/10.1103/PhysRevD.33.3495}
}

@article{Grand:2016mgo,
      author         = "Grand, Robert J. J. and Gómez, Facundo A. and Marinacci,
                        Federico and Pakmor, Ruediger and Springel, Volker and
                        Campbell, David J. R. and Frenk, Carlos S. and Jenkins,
                        Adrian and White, Simon D. M.",
      title          = "{The Auriga Project: the properties and formation
                        mechanisms of disc galaxies across cosmic time}",
      journal        = "Mon. Not. Roy. Astron. Soc.",
      volume         = "467",
      year           = "2017",
      number         = "1",
      pages          = "179-207",
      doi            = "10.1093/mnras/stx071",
      eprint         = "1610.01159",
      archivePrefix  = "arXiv",
      primaryClass   = "astro-ph.GA",
      SLACcitation   = "%%CITATION = ARXIV:1610.01159;%%"
}

@article{Crain:2015poa,
    author = "Crain, Robert A. and others",
    title = "{The EAGLE simulations of galaxy formation: calibration of subgrid physics and model variations}",
    eprint = "1501.01311",
    archivePrefix = "arXiv",
    primaryClass = "astro-ph.GA",
    doi = "10.1093/mnras/stv725",
    journal = "Mon. Not. Roy. Astron. Soc.",
    volume = "450",
    number = "2",
    pages = "1937--1961",
    year = "2015"
}

@article{Schaye:2014tpa,
    author = "Schaye, Joop and others",
    title = "{The EAGLE project: Simulating the evolution and assembly of galaxies and their environments}",
    eprint = "1407.7040",
    archivePrefix = "arXiv",
    primaryClass = "astro-ph.GA",
    doi = "10.1093/mnras/stu2058",
    journal = "Mon. Not. Roy. Astron. Soc.",
    volume = "446",
    pages = "521--554",
    year = "2015"
}

@article{Planck:2015fie,
    author = "Ade, P. A. R. and others",
    collaboration = "Planck",
    title = "{Planck 2015 results. XIII. Cosmological parameters}",
    eprint = "1502.01589",
    archivePrefix = "arXiv",
    primaryClass = "astro-ph.CO",
    doi = "10.1051/0004-6361/201525830",
    journal = "Astron. Astrophys.",
    volume = "594",
    pages = "A13",
    year = "2016"
}

@article{Springel:2009aa,
    author = "Springel, Volker",
    title = "{E pur si muove: Galiliean-invariant cosmological hydrodynamical simulations on a moving mesh}",
    eprint = "0901.4107",
    archivePrefix = "arXiv",
    primaryClass = "astro-ph.CO",
    doi = "10.1111/j.1365-2966.2009.15715.x",
    journal = "Mon. Not. Roy. Astron. Soc.",
    volume = "401",
    pages = "791",
    year = "2010"
}

@ARTICLE{Grand:2024,
       author = {{Grand}, Robert J.~J. and {Fragkoudi}, Francesca and {G{\'o}mez}, Facundo A. and {Jenkins}, Adrian and {Marinacci}, Federico and {Pakmor}, R{\"u}diger and {Springel}, Volker},
        title = "{Overview and public data release of the Auriga Project: cosmological simulations of dwarf and Milky Way-mass galaxies}",
      journal = {arXiv e-prints},
         year = 2024,
        month = jan,
          eid = {arXiv:2401.08750},
        pages = {arXiv:2401.08750},
          doi = {10.48550/arXiv.2401.08750},
archivePrefix = {arXiv},
       eprint = {2401.08750},
 primaryClass = {astro-ph.GA},
       adsurl = {https://ui.adsabs.harvard.edu/abs/2024arXiv240108750G}
}

@article{Power:2002sw,
    author = "Power, Chris and Navarro, J. F. and Jenkins, A. and Frenk, C. S. and White, Simon D. M. and Springel, V. and Stadel, J. and Quinn, Thomas R.",
    title = "{The Inner structure of Lambda CDM halos. 1. A Numerical convergence study}",
    eprint = "astro-ph/0201544",
    archivePrefix = "arXiv",
    doi = "10.1046/j.1365-8711.2003.05925.x",
    journal = "Mon. Not. Roy. Astron. Soc.",
    volume = "338",
    pages = "14--34",
    year = "2003"
}

@ARTICLE{Jenkins2013,
   author = {{Jenkins}, A.},
    title = "{A new way of setting the phases for cosmological multiscale Gaussian initial conditions}",
  journal = {MNRAS},
archivePrefix = "arXiv",
   eprint = {1306.5968},
     year = 2013,
    month = sep,
   volume = 434,
    pages = {2094-2120},
      doi = {10.1093/mnras/stt1154},
   adsurl = {http://adsabs.harvard.edu/abs/2013MNRAS.434.2094J}
}

@article{Bozorgnia:2016ogo,
      author         = "Bozorgnia, Nassim and Calore, Francesca and Schaller,
                        Matthieu and Lovell, Mark and Bertone, Gianfranco and
                        Frenk, Carlos S. and Crain, Robert A. and Navarro, Julio
                        F. and Schaye, Joop and Theuns, Tom",
      title          = "{Simulated Milky Way analogues: implications for dark
                        matter direct searches}",
      journal        = "JCAP",
      volume         = "1605",
      year           = "2016",
      number         = "05",
      pages          = "024",
      doi            = "10.1088/1475-7516/2016/05/024",
      eprint         = "1601.04707",
      archivePrefix  = "arXiv",
      primaryClass   = "astro-ph.CO",
      SLACcitation   = "%%CITATION = ARXIV:1601.04707;%%"
}

@article{Kelso:2016qqj,
      author         = "Kelso, Chris and Savage, Christopher and Valluri, Monica
                        and Freese, Katherine and Stinson, Gregory S. and Bailin,
                        Jeremy",
      title          = "{The impact of baryons on the direct detection of dark
                        matter}",
      journal        = "JCAP",
      volume         = "1608",
      year           = "2016",
      pages          = "071",
      doi            = "10.1088/1475-7516/2016/08/071",
      eprint         = "1601.04725",
      archivePrefix  = "arXiv",
      primaryClass   = "astro-ph.GA",
      reportNumber   = "NORDITA-2015-15, CETUP2015-030",
      SLACcitation   = "%%CITATION = ARXIV:1601.04725;%%"
}

@article{Sloane:2016kyi,
      author         = "Sloane, Jonathan D. and Buckley, Matthew R. and Brooks,
                        Alyson M. and Governato, Fabio",
      title          = "{Assessing Astrophysical Uncertainties in Direct
                        Detection with Galaxy Simulations}",
      journal        = "Astrophys. J.",
      volume         = "831",
      year           = "2016",
      pages          = "93",
      doi            = "10.3847/0004-637X/831/1/93",
      eprint         = "1601.05402",
      archivePrefix  = "arXiv",
      primaryClass   = "astro-ph.GA",
      SLACcitation   = "%%CITATION = ARXIV:1601.05402;%%"
}

@article{Bozorgnia:2017brl,
      author         = "Bozorgnia, Nassim and Bertone, Gianfranco",
      title          = "{Implications of hydrodynamical simulations for the
                        interpretation of direct dark matter searches}",
      journal        = "Int. J. Mod. Phys.",
      volume         = "A32",
      year           = "2017",
      number         = "21",
      pages          = "1730016",
      doi            = "10.1142/S0217751X17300162",
      eprint         = "1705.05853",
      archivePrefix  = "arXiv",
      primaryClass   = "astro-ph.CO",
      SLACcitation   = "%%CITATION = ARXIV:1705.05853;%%"
}

@article{Bozorgnia:2025lsl,
    author = "Bozorgnia, Nassim and Bramante, Joseph and Buchanan, Andrew",
    title = "{High mass dark matter searches with the high speed flux from the Large Magellanic Cloud}",
    eprint = "2511.21841",
    archivePrefix = "arXiv",
    primaryClass = "hep-ph",
    doi = "10.1088/1475-7516/2026/04/065",
    journal = "JCAP",
    volume = "04",
    pages = "065",
    year = "2026"
}

@article{Bozorgnia:2019mjk,
    author = "Bozorgnia, Nassim and Fattahi, Azadeh and Frenk, Carlos S. and Cheek, Andrew and Cerdeno, David G. and G\'omez, Facundo A. and Grand, Robert J.J. and Marinacci, Federico",
    title = "{The dark matter component of the Gaia radially anisotropic substructure}",
    eprint = "1910.07536",
    archivePrefix = "arXiv",
    primaryClass = "astro-ph.GA",
    doi = "10.1088/1475-7516/2020/07/036",
    journal = "JCAP",
    volume = "07",
    pages = "036",
    year = "2020"
}

@article{Kuhlen:2013tra,
    author = "Kuhlen, Michael and Pillepich, Annalisa and Guedes, Javiera and Madau, Piero",
    title = "{The Distribution of Dark Matter in the Milky Way's Disk}",
    eprint = "1308.1703",
    archivePrefix = "arXiv",
    primaryClass = "astro-ph.GA",
    doi = "10.1088/0004-637X/784/2/161",
    journal = "Astrophys. J.",
    volume = "784",
    pages = "161",
    year = "2014"
}

@article{Lawrence:2022niq,
    author = "Lawrence, Grace E. and Duffy, Alan R. and Blake, Chris A. and Hopkins, Philip F.",
    title = "{Gusts in the Headwind: Uncertainties in Direct Dark Matter Detection}",
    eprint = "2207.07644",
    archivePrefix = "arXiv",
    primaryClass = "astro-ph.GA",
    doi = "10.1093/mnras/stac2447",
    month = "7",
    year = "2022"
}

@article{Poole-McKenzie:2020dbo,
    author = "Poole-McKenzie, Robert and Font, Andreea S. and Boxer, Billy and McCarthy, Ian G. and Burdin, Sergey and Stafford, Sam G. and Brown, Shaun T.",
    title = "{Informing dark matter direct detection limits with the ARTEMIS simulations}",
    eprint = "2006.15159",
    archivePrefix = "arXiv",
    primaryClass = "astro-ph.CO",
    doi = "10.1088/1475-7516/2020/11/016",
    journal = "JCAP",
    volume = "11",
    pages = "016",
    year = "2020"
}

@article{Lacroix:2020lhn,
    author = "Lacroix, Thomas and N\'u\~nez-Casti\~neyra, Arturo and Stref, Martin and Lavalle, Julien and Nezri, Emmanuel",
    title = "{Predicting the dark matter velocity distribution in galactic structures: tests against hydrodynamic cosmological simulations}",
    eprint = "2005.03955",
    archivePrefix = "arXiv",
    primaryClass = "astro-ph.GA",
    reportNumber = "LUPM:20-024, IFT-UAM/CSIC-20-64",
    doi = "10.1088/1475-7516/2020/10/031",
    journal = "JCAP",
    volume = "10",
    pages = "031",
    year = "2020"
}

@article{Besla:2019xbx,
      author         = "Besla, Gurtina and Peter, Annika and Garavito-Camargo,
                        Nicolas",
      title          = "{The highest-speed local dark matter particles come from
                        the Large Magellanic Cloud}",
      journal        = "JCAP",
      volume         = "1911",
      year           = "2019",
      number         = "11",
      pages          = "013",
      doi            = "10.1088/1475-7516/2019/11/013",
      eprint         = "1909.04140",
      archivePrefix  = "arXiv",
      primaryClass   = "astro-ph.GA",
      SLACcitation   = "%%CITATION = ARXIV:1909.04140;%%"
}

@article{Donaldson:2021byu,
    author = "Donaldson, Katelin and Petersen, Michael S. and Pe\~narrubia, Jorge",
    title = "{Effects on the local dark matter distribution due to the Large Magellanic Cloud}",
    eprint = "2111.15440",
    archivePrefix = "arXiv",
    primaryClass = "astro-ph.GA",
    doi = "10.1093/mnrasl/slac031",
    journal = "Mon. Not. Roy. Astron. Soc.",
    volume = "L513",
    number = "1",
    pages = "46",
    year = "2022"
}

@article{Lewin:1995rx,
    author = "Lewin, J. D. and Smith, P. F.",
    title = "{Review of mathematics, numerical factors, and corrections for dark matter experiments based on elastic nuclear recoil}",
    doi = "10.1016/S0927-6505(96)00047-3",
    journal = "Astropart. Phys.",
    volume = "6",
    pages = "87--112",
    year = "1996"
}

@article{Reynoso-Cordova:2024lmc,
    author = "Reynoso-Cordova, Javier and Bozorgnia, Nassim and Piro, Marie-C'ecile",
    title = "{The Large Magellanic Cloud: expanding the low-mass parameter space of dark matter direct detection}",
    doi = "10.1088/1475-7516/2024/12/037",
    journal = "JCAP",
    volume = "12",
    pages = "037",
    year = "2024"
}

@article{Mayet:2016zxu,
    author = "Mayet, F. and others",
    title = "{A review of the discovery reach of directional Dark Matter detection}",
    eprint = "1602.03781",
    archivePrefix = "arXiv",
    primaryClass = "astro-ph.CO",
    doi = "10.1016/j.physrep.2016.02.007",
    journal = "Phys. Rept.",
    volume = "627",
    pages = "1--49",
    year = "2016"
}

@article{Baracchini:2023kyk,
    author = "Baracchini, Elisabetta",
    title = "{Directional dark matter searches}",
    doi = "10.21468/SciPostPhysProc.12.002",
    journal = "SciPost Phys. Proc.",
    volume = "12",
    pages = "002",
    year = "2023"
}

@article{Vahsen:2021gnb,
    author = "Vahsen, Sven E. and O'Hare, Ciaran A. J. and Loomba, Dinesh",
    title = "{Directional Recoil Detection}",
    eprint = "2102.04596",
    archivePrefix = "arXiv",
    primaryClass = "physics.ins-det",
    doi = "10.1146/annurev-nucl-020821-035016",
    journal = "Ann. Rev. Nucl. Part. Sci.",
    volume = "71",
    pages = "189--224",
    year = "2021"
}

@article{OHare:2021utq,
    author = "O'Hare, Ciaran A. J.",
    title = "{New Definition of the Neutrino Floor for Direct Dark Matter Searches}",
    eprint = "2109.03116",
    archivePrefix = "arXiv",
    primaryClass = "hep-ph",
    doi = "10.1103/PhysRevLett.127.251802",
    journal = "Phys. Rev. Lett.",
    volume = "127",
    number = "25",
    pages = "251802",
    year = "2021"
}

@article{Spergel:1987kx,
    author = "Spergel, David N.",
    title = "{The Motion of the Earth and the Detection of Wimps}",
    reportNumber = "IASSNS-AST-87/2",
    doi = "10.1103/PhysRevD.37.1353",
    journal = "Phys. Rev. D",
    volume = "37",
    pages = "1353",
    year = "1988"
}

@article{Snowden-Ifft:1999reu,
    author = "Snowden-Ifft, D. P. and Martoff, C. J. and Burwell, J. M.",
    title = "{Low pressure negative ion drift chamber for dark matter search}",
    eprint = "astro-ph/9904064",
    archivePrefix = "arXiv",
    doi = "10.1103/PhysRevD.61.101301",
    journal = "Phys. Rev. D",
    volume = "61",
    pages = "101301",
    year = "2000"
}

@article{Burgos:2007zz,
    author = "Burgos, S. and others",
    title = "{First results from the DRIFT-IIa dark matter detector}",
    eprint = "0707.1488",
    archivePrefix = "arXiv",
    primaryClass = "hep-ex",
    doi = "10.1016/j.astropartphys.2007.08.007",
    journal = "Astropart. Phys.",
    volume = "28",
    pages = "409--421",
    year = "2007"
}

@article{Billard_2012,
doi = {10.1088/1742-6596/375/1/012008},
url = {https://doi.org/10.1088/1742-6596/375/1/012008},
year = {2012},
month = {jul},
publisher = {},
volume = {375},
number = {1},
pages = {012008},
author = {Billard, J and Mayet, F and Santos, D},
title = {Directional Detection of Dark Matter with MIMAC},
journal = {Journal of Physics: Conference Series}
}

@article{Santos_2018,
doi = {10.1088/1742-6596/1029/1/012005},
url = {https://doi.org/10.1088/1742-6596/1029/1/012005},
year = {2018},
month = {may},
publisher = {IOP Publishing},
volume = {1029},
number = {1},
pages = {012005},
author = {Santos, Daniel},
title = {Dark Matter Directional Detection with MIMAC},
journal = {Journal of Physics: Conference Series}
}

@article{Amaro:2022gub,
    author = "Amaro, Fernando Domingues and others",
    title = "{The CYGNO Experiment}",
    eprint = "2202.05480",
    archivePrefix = "arXiv",
    primaryClass = "physics.ins-det",
    doi = "10.3390/instruments6010006",
    journal = "Instruments",
    volume = "6",
    number = "1",
    pages = "6",
    year = "2022"
}

@article{Ahlen:2009ev,
    author = "Ahlen, S. and others",
    title = "{The case for a directional dark matter detector and the status of current experimental efforts}",
    eprint = "0911.0323",
    archivePrefix = "arXiv",
    primaryClass = "astro-ph.CO",
    doi = "10.1142/S0217751X10048172",
    journal = "Int. J. Mod. Phys. A",
    volume = "25",
    pages = "1--51",
    year = "2010"
}

@article{Ahlen:2010ub,
    author = "Ahlen, S. and others",
    title = "{First Dark Matter Search Results from a Surface Run of the 10-L DMTPC Directional Dark Matter Detector}",
    eprint = "1006.2928",
    archivePrefix = "arXiv",
    primaryClass = "hep-ex",
    doi = "10.1016/j.physletb.2010.11.041",
    journal = "Phys. Lett. B",
    volume = "695",
    pages = "124--129",
    year = "2011"
}

@article{Miuchi:2010hn,
    author = "Miuchi, Kentaro and others",
    title = "{First underground results with NEWAGE-0.3a direction-sensitive dark matter detector}",
    eprint = "1002.1794",
    archivePrefix = "arXiv",
    primaryClass = "astro-ph.CO",
    doi = "10.1016/j.physletb.2010.02.028",
    journal = "Phys. Lett. B",
    volume = "686",
    pages = "11--17",
    year = "2010"
}

@article{Shimada:2023vky,
    author = "Shimada, Takuya and others",
    title = "{Direction-sensitive dark matter search with 3D-vector-type tracking in NEWAGE}",
    eprint = "2301.04779",
    archivePrefix = "arXiv",
    primaryClass = "hep-ex",
    doi = "10.1093/ptep/ptad120",
    journal = "PTEP",
    volume = "2023",
    number = "10",
    pages = "103F01",
    year = "2023"
}

@article{Nakamura:2015iza,
    author = "Nakamura, Kiseki and others",
    title = "{Direction-sensitive dark matter search with gaseous tracking detector NEWAGE-0.3b{\textquoteright}}",
    doi = "10.1093/ptep/ptv041",
    journal = "PTEP",
    volume = "2015",
    number = "4",
    pages = "043F01",
    year = "2015"
}

@inproceedings{Alexandrov:2019gme,
    author = "Alexandrov, Andrey",
    collaboration = "NEWSdm",
    title = "{Directional Detection of Dark Matter With a Nuclear Emulsion Based Detector}",
    booktitle = "{18th Lomonosov Conference on Elementary Particle Physics}",
    doi = "10.1142/9789811202339_0060",
    pages = "314--317",
    year = "2019"
}

@article{Golovatiuk:2021jcf,
    author = "Golovatiuk, Artem",
    collaboration = "NEWSdm",
    title = "{Directional Dark Matter Search with the NEWSdm experiment}",
    doi = "10.1088/1742-6596/2156/1/012044",
    journal = "J. Phys. Conf. Ser.",
    volume = "2156",
    pages = "012044",
    year = "2021"
}

@phdthesis{Schuster:2016kpt,
    author = "Schuster, Patricia Frances",
    title = "{Investigating the Anisotropic Scintillation Response in Organic Crystal Scintillator Detectors}",
    school = "UC, Berkeley (main)",
    year = "2016"
}

@article{SHIMIZU2003347,
title = {Directional scintillation detector for the detection of the wind of WIMPs},
journal = {Nuclear Instruments and Methods in Physics Research Section A: Accelerators, Spectrometers, Detectors and Associated Equipment},
volume = {496},
number = {2},
pages = {347-352},
year = {2003},
issn = {0168-9002},
doi = {https://doi.org/10.1016/S0168-9002(02)01661-3},
url = {https://www.sciencedirect.com/science/article/pii/S0168900202016613},
author = {Y Shimizu and M Minowa and H Sekiya and Y Inoue}
}

@article{Nygren:2013nda,
    author = "Nygren, D. R.",
    editor = "Irastorza, Igor G. and Colas, Paul and Giomataris, Iioannis",
    title = "{Columnar recombination: a tool for nuclear recoil directional sensitivity in a xenon-based direct detection WIMP search}",
    doi = "10.1088/1742-6596/460/1/012006",
    journal = "J. Phys. Conf. Ser.",
    volume = "460",
    pages = "012006",
    year = "2013"
}

@article{Mohlabeng:2015efa,
    author = "Mohlabeng, Gopolang and Kong, Kyoungchul and Li, Jin and Para, Adam and Yoo, Jonghee",
    title = "{Dark Matter Directionality Revisited with a High Pressure Xenon Gas Detector}",
    eprint = "1503.03937",
    archivePrefix = "arXiv",
    primaryClass = "hep-ph",
    reportNumber = "FERMILAB-PUB-15-079-E",
    doi = "10.1007/JHEP07(2015)092",
    journal = "JHEP",
    volume = "07",
    pages = "092",
    year = "2015"
}

@article{Baracchini:2020btb,
    author = "Baracchini, E. and others",
    title = "{CYGNO: a gaseous TPC with optical readout for dark matter directional search}",
    eprint = "2007.12627",
    archivePrefix = "arXiv",
    primaryClass = "physics.ins-det",
    doi = "10.1088/1748-0221/15/07/C07036",
    journal = "JINST",
    volume = "15",
    number = "07",
    pages = "C07036",
    year = "2020"
}

@article{Vahsen:2020pzb,
    author = "Vahsen, S. E. and others",
    title = "{CYGNUS: Feasibility of a nuclear recoil observatory with directional sensitivity to dark matter and neutrinos}",
    eprint = "2008.12587",
    archivePrefix = "arXiv",
    primaryClass = "physics.ins-det",
    month = "8",
    year = "2020"
}

@article{Lisotti:2024fco,
    author = "Lisotti, Chiara and others",
    title = "{CYG$\nu $S: detecting solar neutrinos with directional gas time projection chambers}",
    eprint = "2404.03690",
    archivePrefix = "arXiv",
    primaryClass = "hep-ph",
    doi = "10.1140/epjc/s10052-024-13392-3",
    journal = "Eur. Phys. J. C",
    volume = "84",
    number = "10",
    pages = "1021",
    year = "2024"
}

@techreport{Mazzitelli:2023tdr,
      author       = "Baracchini, E. and others",
      title        = "{Technical Design Report — TDR CYGNO-04/INITIUM}",
      institution  = "INFN",
      number       = "INFN-23-06-LNF",
      year         = "2023",
      url          = "https://doi.org/10.15161/oar.it/76967"
}

@article{LZ:2024zvo,
    author = "Aalbers, J. and others",
    collaboration = "LZ",
    title = "{Dark Matter Search Results from 4.2{\,}{\,}Tonne-Years of Exposure of the LUX-ZEPLIN (LZ) Experiment}",
    eprint = "2410.17036",
    archivePrefix = "arXiv",
    primaryClass = "hep-ex",
    reportNumber = "FERMILAB-PUB-24-0796-V",
    doi = "10.1103/4dyc-z8zf",
    journal = "Phys. Rev. Lett.",
    volume = "135",
    number = "1",
    pages = "011802",
    year = "2025"
}

@article{Baxter:2021pqo,
    author = "Baxter, D. and others",
    title = "{Recommended conventions for reporting results from direct dark matter searches}",
    eprint = "2105.00599",
    archivePrefix = "arXiv",
    primaryClass = "hep-ex",
    doi = "10.1140/epjc/s10052-021-09655-y",
    journal = "Eur. Phys. J. C",
    volume = "81",
    number = "10",
    pages = "907",
    year = "2021"
}

@article{Gorski:2004by,
    author = "G{\'o}rski, K. M. and Hivon, E. and Banday, A. J. and Wandelt, B. D. and Hansen, F. K. and Reinecke, M. and Bartelman, M.",
    title = "{HEALPix - A Framework for high resolution discretization, and fast analysis of data distributed on the sphere}",
    eprint = "astro-ph/0409513",
    archivePrefix = "arXiv",
    doi = "10.1086/427976",
    journal = "Astrophys. J.",
    volume = "622",
    pages = "759--771",
    year = "2005"
}

@article{Coquillat:2025ahd,
    author = "Coquillat, Jean-Marie",
    title = "{Light WIMP search with the NEWS-G experiment}",
    doi = "10.22323/1.501.0477",
    journal = "PoS",
    volume = "ICRC2025",
    pages = "477",
    year = "2025"
}

\end{document}